# Hypervelocity Impact Debris Cloud Trajectory-Planning based on Additive Manufactured Lattice Structures

**Bilin Zheng[1], Xiao Kang[1*], Xiaoyu Zhang[1,2*], Hao Zhou[2],Mengchuan Xu[2],Chang Liu[3]**

[1] Institute of Advanced Structure Technology, Beijing Institute of Technology, Beijing 100081, China
[2] Beijing Key Laboratory of Intelligent Space Robotic Systems Technology and Applications, Beijing Institute of Spacecraft System Engineering, China Academy of Space Technology, Beijing 100094, China
[3]Dalian University of Technology，Dalian 116024，China
**\* Correspondence:**
Corresponding Author:
Xiao Kang, xiao.kang@bit.edu.cn
Xiaoyu Zhang, 08121956@bjtu.edu.cn

Abstract

Space debris and micrometeoroid (MMOD) impacts pose a serious threat to the safe operation of spacecraft. However, traditional protective structures typically suffer from limitations such as excessive thickness and inadequate load-bearing capacity. Guided by the design concepts of debris-cloud deflection and hierarchical energy dissipation, this study proposes a trajectory-planning lattice protective structure. First, the lattice parameters and geometry were designed according to the functional relationship between the incident angle and the transmitted/ricochet trajectory angles. Subsequently, multi-angle hypervelocity impact experiments were carried out to evaluate the proposed lattice protection structure. In combination with post-impact CT three-dimensional reconstruction and smoothed particle hydrodynamics (SPH) numerical simulations, the protective mechanisms of the lattice structure were systematically characterized and clarified. The results demonstrate that, for three oblique incidence conditions, the lattice structure remained intact and significantly deflected the debris-cloud momentum direction while effectively dissipating its kinetic energy. The experimental trajectories were found to closely match the theoretical predictions. The angled plates with gradient designs enabled continuous changes in the momentum direction and stepwise kinetic energy dissipation through multiple cycles of debrisation, dispersion, and trajectory deflection. As the number of penetrated layers increased, the energy-dissipation mechanism transitioned from a hypervelocity regime dominated by phase transformation and brittle fracture to one primarily governed by plastic deformation. The ballistic limit curve for this structure was obtained; compared with traditional aluminum-alloy Whipple shields, the total thickness was reduced by approximately 25%, while the maximum protected projectile diameter increased by an average of 29.1%. This research presents a novel, engineering-ready approach for spacecraft MMOD protection and validates the potential of trajectory-planning lattice structures for hypervelocity impact defense.

## 1 Introduction

Micrometeoroids and space debris in Earth orbit pose substantial risks to spacecraft operations because of their hypervelocity impacts[1-4]. Such debris typically collides with spacecraft at velocities exceeding 3 km/s, generating impact pressures that far exceed structural strength limits—characteristic of hypervelocity impact phenomena[5-7]. Since 1957, accumulated space debris has increasingly saturated near-Earth and geostationary orbits, with the total number of objects exceeding 100 million[8, 9]. The vast quantity of hypervelocity orbital debris poses a significant threat to spacecraft[10, 11]. The current mass of space debris in Earth orbit is estimated to be approximately 9,000 tons[12]. According to projections by the European Space Agency, the amount of space debris is expected to increase by approximately 10% annually over the next 50 years, with the highest concentration predicted in the 700–1,100 km low-Earth-orbit region, which is densely populated with remote-sensing and mobile-communication satellites[13, 14]. Due to the continuous growth of space debris, impacts between debris and spacecraft have become increasingly frequent in recent years. For instance, in March 2021, the

Yunhai-02 satellite disintegrated following an impact, and in May 2022, the James Webb Space Telescope sustained damage to its mirror assembly as a result of an asteroid impact. As space activities continue to expand, it is expected that the threat posed by space debris impacts will persist and intensify over the long term. In addition to developing active defense measures such as deceleration and evasion, passive protection is equally essential for the long-term safeguarding of spacecraft[15, 16]. Such passive protection can effectively mitigate damage caused by numerous micrometeoroid impacts[17, 18]. Therefore, designing effective protective structures to shield spacecraft from hypervelocity space debris impacts constitutes a critical challenge in aerospace engineering.

Currently, the primary protective architecture for large crewed spacecraft, such as the International Space Station, against hypervelocity impacts is the Whipple shield, which consists of multiple material layers[2, 19]. The Whipple shield was first proposed in 1947 by the American astronomer Fred Whipple[16, 20]. Its fundamental design involves positioning a thin bumper at a prescribed distance from the spacecraft's outer wall panel. Upon hypervelocity collision with debris, this bumper causes the projectile to shatter, melt, or vaporize, thereby generating a high-speed debris cloud. This cloud then expands and disperses within the gap between the two wall panels, thereby dissipating the original projectile's momentum and kinetic energy[18, 21]. After undergoing further debrisation and spatial dispersion, the resulting debris cloud ultimately impacts the rear primary wall panel with a significantly reduced energy flux per unit area, thereby providing effective protection for the spacecraft's main structure[22, 23]. However, the Whipple configuration requires a relatively large interlayer spacing to fully exploit the debris cloud dispersion mechanism[24]. The traditional Whipple protective structure itself has limited load-bearing capacity, which makes such designs difficult to apply to small spacecraft such as satellites. Most existing research on hypervelocity impact protection structures centers on Whipple-type designs, focusing on enhancing protective materials, optimizing parameters such as panel thickness and spacing, or employing composite and functionally graded materials to improve performance[25,

26]. Consequently, innovation in structural concepts remains limited, necessitating the exploration of novel protective configurations to overcome the constraints of traditional Whipple structures in terms of interlayer spacing, mass efficiency, and load-bearing capacity.

In recent years, structural innovations for hypervelocity impact protection have attracted increasing attention[27-31]. Some researchers have attempted to replace traditional protective layers with simple lattice-type structures such as corrugated plates or honeycomb sandwich panels, aiming to provide load-bearing capacity while simultaneously dispersing debris clouds. However, the honeycomb cells in typical honeycomb sandwich structures exhibit a so-called "channelling effect," in which debris travels directly through the honeycomb channels at high speed to impact the rear panel, thereby reducing protective effectiveness[32-34]. Corrugated sandwich structures offer only marginally better protection than multilayer Whipple structures of equivalent total thickness. Existing structural innovations remain limited and generally yield only incremental improvements in protective performance[35, 36]. Moreover, additive manufacturing (AM) technology has gained attention for the development of novel hypervelocity impact protection structures due to its ability to fabricate lightweight, highly complex geometries that are not achievable with conventional manufacturing. Bruce A. Davis[37] pioneered the development of AlSi10Mg body-centered cubic (BCC) and Kelvin lattice protective structures fabricated by AM in 2019, experimentally demonstrating the engineering feasibility of AM-based aluminum alloy hypervelocity impact protection concepts. Feier et al.[38] designed multiple novel protective structures using high-performance thermoplastics processed by AM to disperse debris clouds and demonstrated their effectiveness in secondary light-gun tests. Olivieri et al.[39] developed corrugated-plate protective structures for microsatellites using AM. Hypervelocity impact test results demonstrated superior protective performance and load-bearing capacity compared with traditional honeycomb sandwich panels. These studies collectively indicate the significant potential of AM lattice structures for hypervelocity debris protection[40].

To address the limitations of existing protective structures, this study proposes a lattice protection structure based on additive manufacturing that controls debris cloud trajectories. Departing from traditional protective concepts, this structure employs a multi-level combination of inclined plate units to construct a lattice. This design guides hypervelocity debris clouds through continuous deflection during penetration of the protective layer, progressively dissipating their kinetic energy. This research aims to propose a more effective novel lattice protection structure for spacecraft against hypervelocity impacts, providing a new technical pathway for the design of spacecraft protection structures.

## 2 Debris cloud trajectory-planning lattice structure

The lattice structure-type space debris shield proposed in this study comprises a bumper, a set of multi-tier inclined plates, and a rear wall, as shown in Figure 1. The bumper serves to debris the projectile upon impact, thereby generating a debris cloud. This cloud is subsequently guided through multiple inclined plates arranged at different angles, causing the hypervelocity debris cloud to undergo sequential deflections as it penetrates the protective layers. This process reduces the normal component of the debris cloud's kinetic energy and redistributes its total kinetic energy over a larger spatial region. To ensure consistent protective performance regardless of the impact location on the lattice shield, an axially symmetric configuration is adopted. The following sections present the fundamental design methodology of the lattice shield.

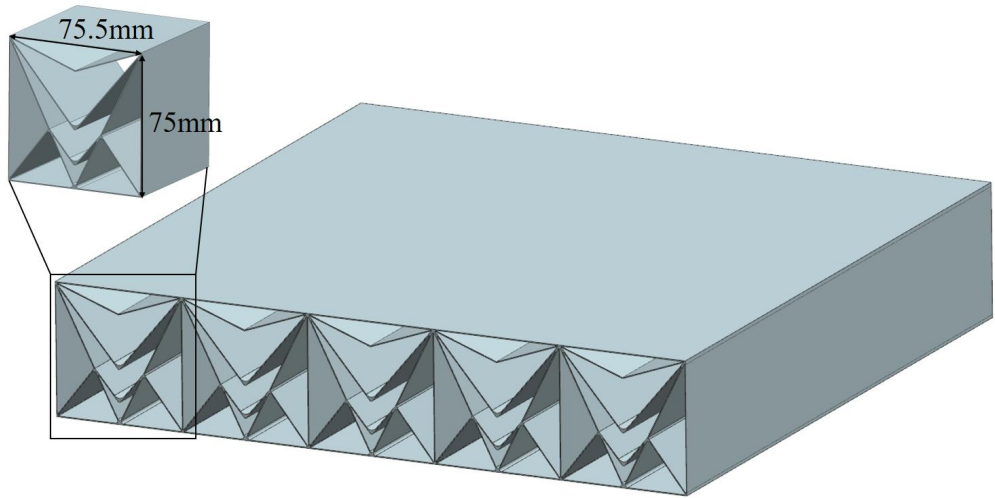

Figure 1. Lattice shield model.

William P. Schonberg[41] conducted oblique impact tests in which projectile velocities ranged from 5 km/s to 8 km/s and impact angles varied between 30° and 75°. Post-test observations of the impact craters on the test plate revealed that, when the debris cloud strikes at oblique angles between 30° and 75°, it splits into three distinct components: the axial debris cloud, the normal debris cloud, and the ricochet debris cloud, as illustrated in Figure 2[42]. In Figure 2, the incident angle is denoted by $\alpha$, the axial trajectory angle by $\theta_1$, the normal trajectory angle by $\theta_2$, and the ricochet debris cloud trajectory angle by $\alpha_c$. The axial debris cloud primarily consists of projectile debris, which exhibits higher velocities and forms larger craters on the observation plate. The normal debris cloud predominantly comprises bumper debris with comparatively lower velocities. The ricochet debris cloud is emitted from above the target plate, with its kinetic energy gradually increasing as the incident angle increases. Based on William P. Schonberg's oblique impact test conditions, numerical simulations using the smoothed particle hydrodynamics (SPH) method were performed under the same conditions. In Figure 3, red represents projectile debris, while blue indicates target plate debris. The numerical simulation results in Figure 3 clearly reproduce the experimental findings: at an incidence angle of 45°, the mass fraction of the ricochet debris cloud increases markedly compared to smaller angles; when the angle of incidence exceeds 60°, the projectile debris are divided into ricochet debris clouds and axial debris clouds, each accounting for approximately half of the total projectile-debris mass; and when the angle of incidence reaches 75°, nearly all projectile debris are ejected as ricochet debris. Since the kinetic energy of the normal debris cloud is smaller than that of the axial debris cloud, only the axial debris cloud is considered when evaluating the transmitted debris cloud. For descriptive convenience, the axial debris cloud is hereafter referred to as the transmitted debris cloud in this paper. Both experimental data from the literature and the numerical simulation results of this work indicate that the transmission angle and ricochet angle of the axial debris cloud exhibit a clear functional dependence on the angle of incidence, as shown in Figure 4.

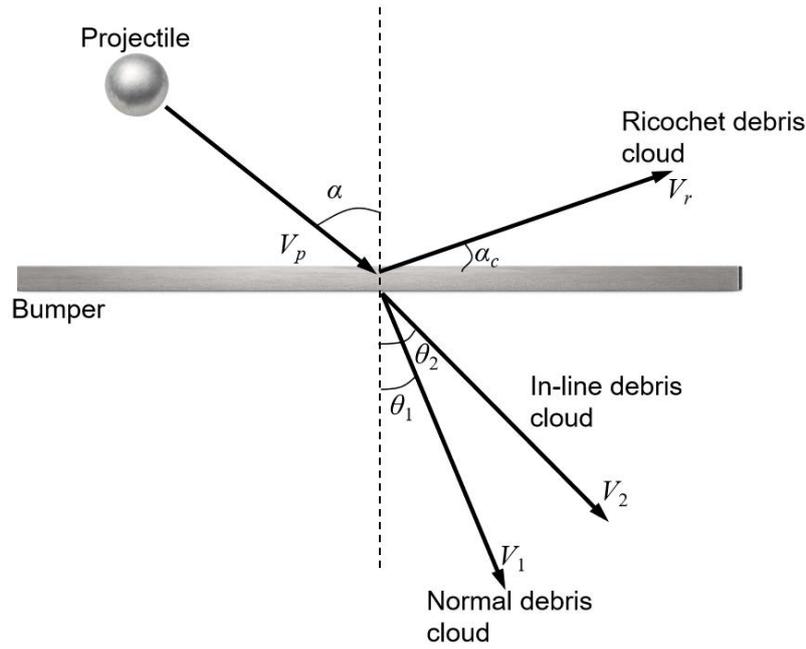

Figure 2. Schematic of the debris cloud generated by an oblique impact.

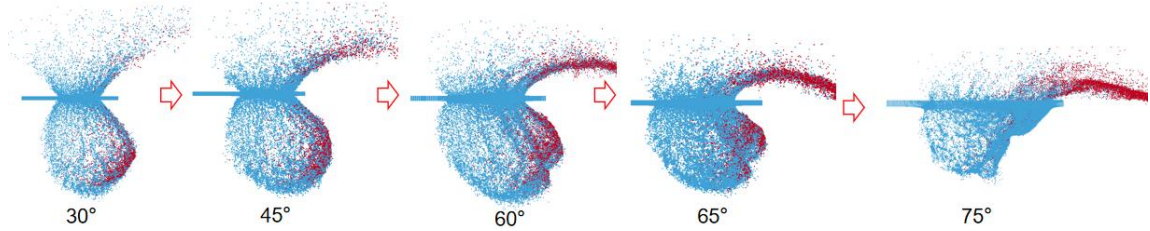

Figure 3. SPH numerical simulation of a hypervelocity oblique impact.

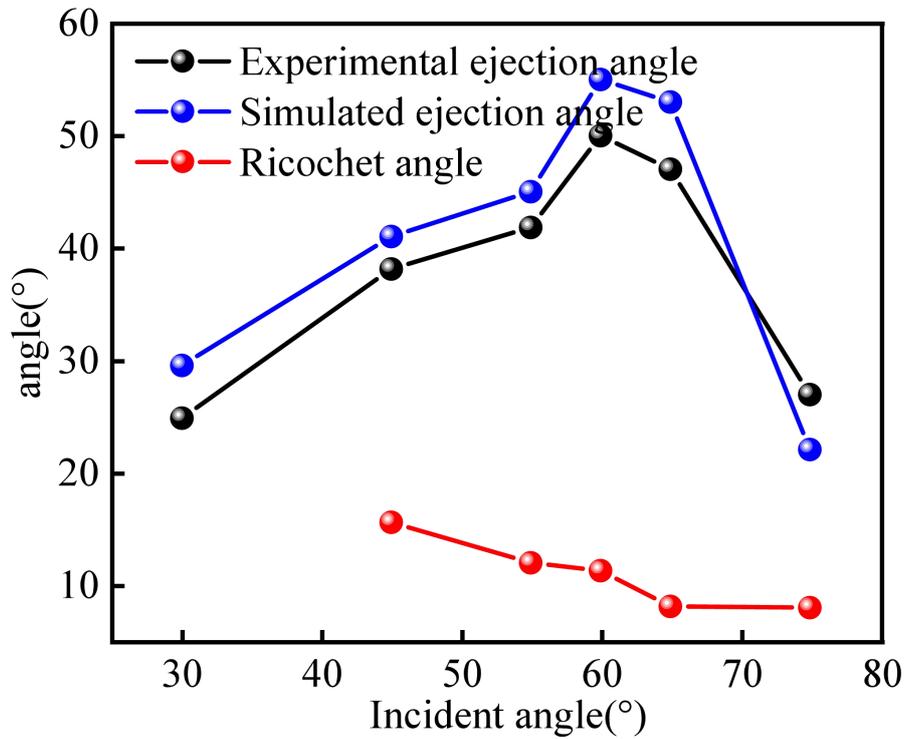

Figure 4. Experimental and numerical data for the relationship between transmission

angle, ricochet angle, and incidence angle.

By combining the experimental data and numerical simulation results with dimensional analysis, the functional relationships between the incident angle and both the axial and ricochet debris cloud trajectory angles are derived. These analyses account for the direct influence of the incident velocity, target-plate thickness, and projectile characteristics, and the resulting expressions are nondimensionalized[41]：

$$\gamma_1/\theta = 0.417(V/C)^{0.228} cos^{0.225}\theta(t_s/d)^{-0.491} \quad (1)$$

$$\alpha_c/\theta = 0.033(V/C)^{0.982} cos^{-3.215}\theta(t_s/d)^{-0.531} \quad (2)$$

Equations (1) and (2) provide the transmission and ricochet angles for projectiles impacting at various oblique angles. Therefore, when a debris cloud is incident vertically, its trajectory can be controlled by a multi-tier inclined-plate structure designed with different inclination angles. This configuration causes the debris cloud's kinetic energy to be repeatedly dispersed upon hypervelocity impact with the inclined plates, while simultaneously deflecting the transmitted and ricochet directions of the debris cloud to be as close as possible to perpendicular to the incident direction. This design substantially reduces the debris cloud's normal kinetic energy relative to the target plate, thereby minimizing damage to the plate. Figure 5(a) illustrates the trajectory of the debris cloud after successive collisions with inclined plates of different inclination angles. From an engineering perspective, the shield must provide consistent protection regardless of the impact location on its surface. Therefore, the lattice shield is designed to be symmetric, with identical upper and lower plate lengths. By extending and combining the inclined plates in Figure 5(a) according to this principle, the lattice shield configuration shown in Figure 5(b) is obtained. The resulting lattice protective structure comprises a bumper, five inclined plates, side plates, and a rear wall. The bumper shatters incoming projectiles into a debris cloud, initially dissipating a portion of their kinetic energy. The five inclined plates further disperse and deflect the debris cloud. The side plates can be regarded as special inclined plates with a 90 ° inclination angle, and they protect against debris clouds transmitted beyond the second inclined plate. The third inclined plate extends to the bottom of the structure and is connected to the rear wall. The third inclined plate and the side plates constitute the primary load-bearing members of the lattice structure under impact loading.

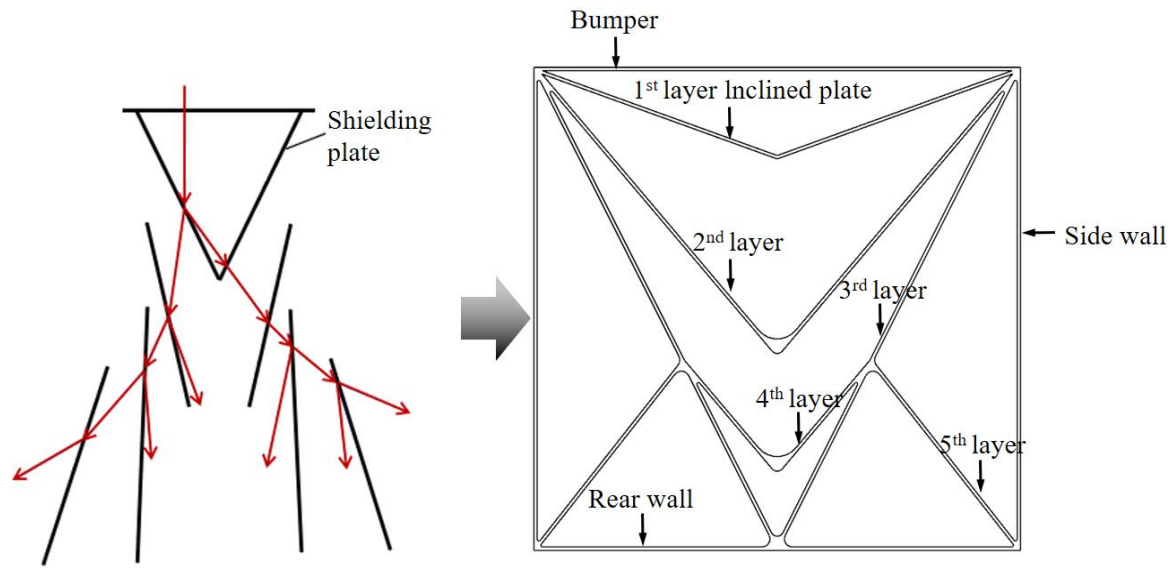

Figure 5. Design of the lattice structure.

Based on Equations (1) and (2), the trajectories of the debris clouds within the structure can be derived for projectiles impacting the lattice shield at different incidence angles. Red arrows indicate the trajectories of transmitted debris clouds, whereas green arrows denote the trajectories of ricochet debris clouds. Figure 6 presents schematic diagrams of the debris cloud trajectories for three representative incident angles. As discussed above, significant ricochet debris clouds are generated only when the incident angle α is greater than or equal to 45°. Therefore, in the trajectory analysis conducted in this study, when the collision angle between the debris and the protective plate is less than 45°, ricochet debris clouds are neglected. Conversely, when α ≥ 75°, transmitted debris is virtually absent, and the debris is ejected predominantly as ricochet debris. Based on the above theoretical derivation, the debris cloud trajectories corresponding to three representative incidence angles are described as follows. As shown in Figure 6(a), when the projectile strikes at 0° (normal incidence), it is shattered by the bumper. The angle between the first inclined plate and the debris is only 20°, and thus, no ricochet debris is produced. The debris then strikes the second layer of inclined plates at an angle of 50°, producing a ricochet debris cloud as it is deflected. The collision process with the third layer of inclined plates is similar to that with the second layer. After colliding with the side plates, the debris ultimately deviates by 45° from the incident direction. As shown in Figure 6(b), when the projectile strikes at 30°, debris collide with the first inclined plate at 50°, generating both a ricochet debris cloud and a transmitted debris cloud. The ricochet debris cloud undergoes secondary ricochets after collision, whereas the transmitted debris cloud experiences only slight deflection and does not generate additional ricochet debris because of the low-angle collisions with the lower inclined plates. Thus, an incident angle of 30° represents a critical condition at which the protective efficacy may be compromised; this conclusion will be validated by subsequent experiments and numerical simulations. At an incidence angle of 60°, ricochet occurs at the bumper, and the normal component of the debris cloud's kinetic energy is already low. When the debris then collide with the first inclined plate at an angle of 70°, most debris ricochet, and the transmitted debris possesses low kinetic energy. After two ricochet events, the debris cloud disperses into two parts: one part is ejected toward the bumper, and the other part continues to be transmitted, as shown in Figure 6(c).

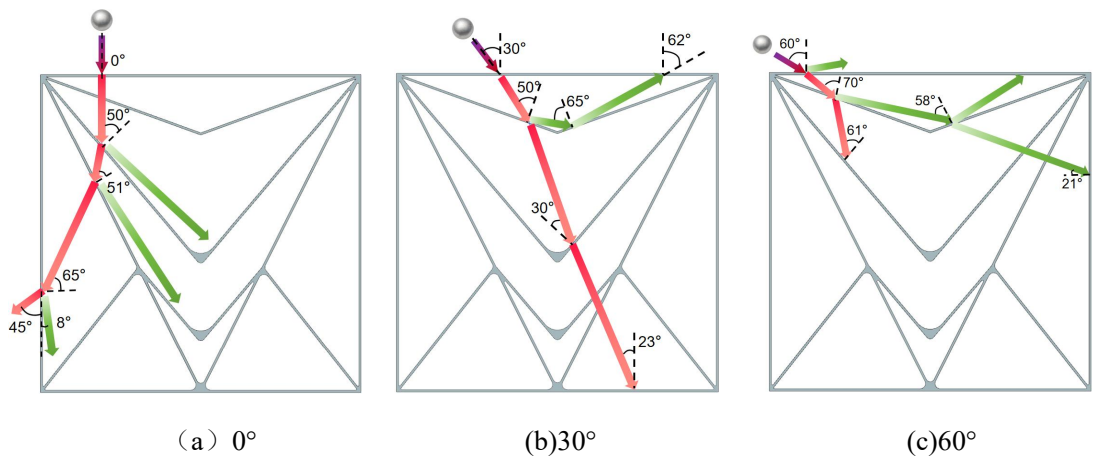

(a) 0°    (b) 30°    (c) 60°

Figure 6. Theoretical debris-cloud trajectories in the lattice shield derived from the trajectory equations.

The three-dimensional schematic of the lattice shield is shown in Figure 6. Each unit cell has a length of 75.5 mm, and the overall shield thickness is 75 mm. This corresponds to a 25.0–37.5% reduction relative to the 100–120 mm total thickness of conventional Whipple structures. This significant reduction in thickness enables the lattice shield to be applied to small spacecraft, including high-value satellites. Due to the structural complexity of the lattice shield, it was fabricated using Selective Laser Melting (SLM). As an advanced additive manufacturing technique, SLM constructs intricate three-dimensional geometries by sequentially melting metal powder layer by layer, providing high precision and excellent control over material properties. The material employed was an AlSi10Mg alloy. The AlSi10Mg alloy powder was produced via gas atomization, with particle sizes ranging from 15μm to 53μm. Its chemical composition comprises aluminum (approximately 87.9 wt.%), silicon (approximately 10.1 wt.%), and magnesium (approximately 0.40 wt.%). Fabrication was performed using an EOS M290 SLM system with a scanning strategy that featured a 67° rotation between successive layers, and the process parameters are listed in Table 1. All SLM printing processes were conducted under an argon atmosphere, thereby ensuring material purity and stable mechanical properties during fabrication. Heat treatment was conducted at 260 °C with a soak time of 2 h, followed by air cooling. The porosity of additively manufactured metals can significantly influence their mechanical properties under hypervelocity impact. To characterize and minimize this effect, high-precision industrial X-ray computed tomography (CT) was employed to scan samples fabricated under the same process parameters as the lattice shield. CT scanning was performed using a μ-CT system (Beijing Institute of Technology, China), with the source voltage and current set to 100 kV and 80 mA, respectively. The pixel size of the tomographic images was 5.09 μm. The commercial software Avizo 2022 was employed for image denoising, segmentation, three-dimensional visualization, and porosity-defect quantification. This analysis yielded quantitative data on the morphology and size of porosity defects, along with three-dimensional models of samples containing actual defects. In total, 10 samples underwent CT scanning. The porosity of each sample is listed in Table 1, with an average value of only 0.00041%. The samples printed using the process employed in this study thus exhibit high density. Figure 7 shows a three-dimensional reconstructed model and the pore-defect distribution of a typical sample. It can be observed that the pores are small and sparsely distributed throughout the sample volume. The extremely low pore content has a negligible effect on the dynamic response of the material under hypervelocity impact.

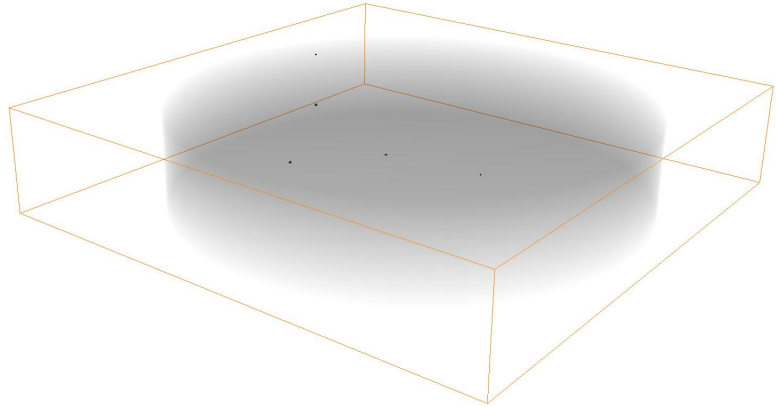

Figure 7. CT reconstruction and pore distribution of a representative AlSi10Mg specimen.

## 3 Hypervelocity impact experiment

The objective of this work is to design a more effective spacecraft shielding system by planning the trajectories of debris clouds. To validate the protective performance of the proposed lattice shield under hypervelocity impact conditions, three hypervelocity impact experiments were conducted. Since the protective concept of the lattice shield relies on controlling the debris-cloud direction through a graded arrangement of inclined shield plates, experimental verification was required to assess the shield's effectiveness against projectiles impacting at different incidence angles.

### 3.1 Experimental Methods

The lattice shield employed in the hypervelocity impact experiments is shown in Figure 7. The test specimen has overall dimensions of 200 × 200 × 75 mm. To ensure a stable bolted assembly to the target fixture during testing, the regions surrounding the bolt holes at the four corners of the structure were printed as solid sections. The hypervelocity impact experiments were conducted using a two-stage light-gas gun over a velocity range of 3–8 km/s. The experimental principle is illustrated schematically in Figure 8. Figure 9 presents a schematic diagram of the hypervelocity impact test system. The experiments were conducted using an 18 mm-bore

hydrogen–oxygen detonation two-stage light-gas gun, which is capable of accelerating projectiles with diameters from 2 to 12 mm to velocities between 2 and 8 km/s. Projectile separation from the baseplate occurred via aerodynamic forces within an air-filled chamber, and the projectile velocity was measured using a laser chronograph. A high-speed camera was used to record the projectile impact on the target plate at a frame rate of 1 million frames per second. High-power LEDs were positioned near the target-chamber window to obtain clear high-speed images at short exposure times. These images enabled the reconstruction of the debris-cloud trajectories within the lattice structure. Hypervelocity imaging was also used to determine the velocity of the debris cloud, with an error of ±5% compared to the projectile velocity measured by the laser chronograph.

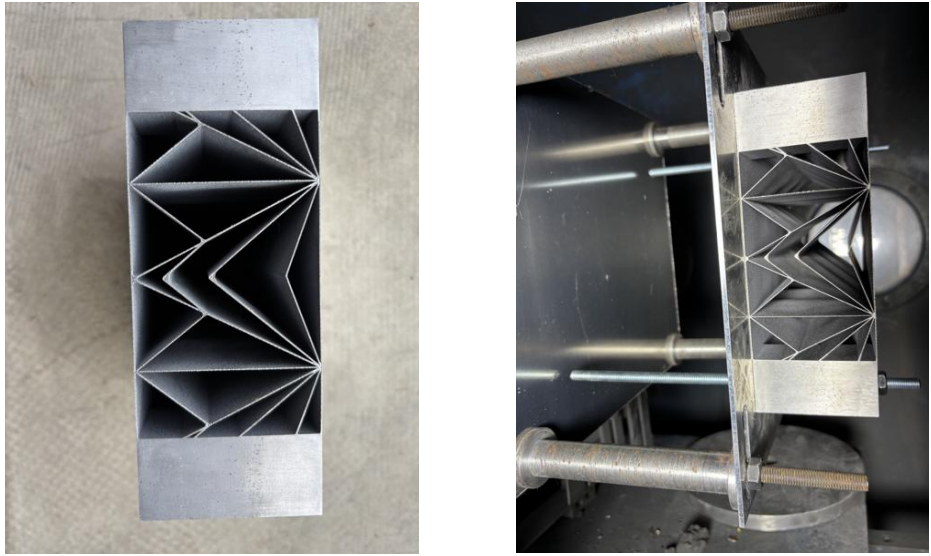

Figure 8. Lattice shield: (a) as-fabricated specimen; (b) lattice shield mounted on the experimental target frame.

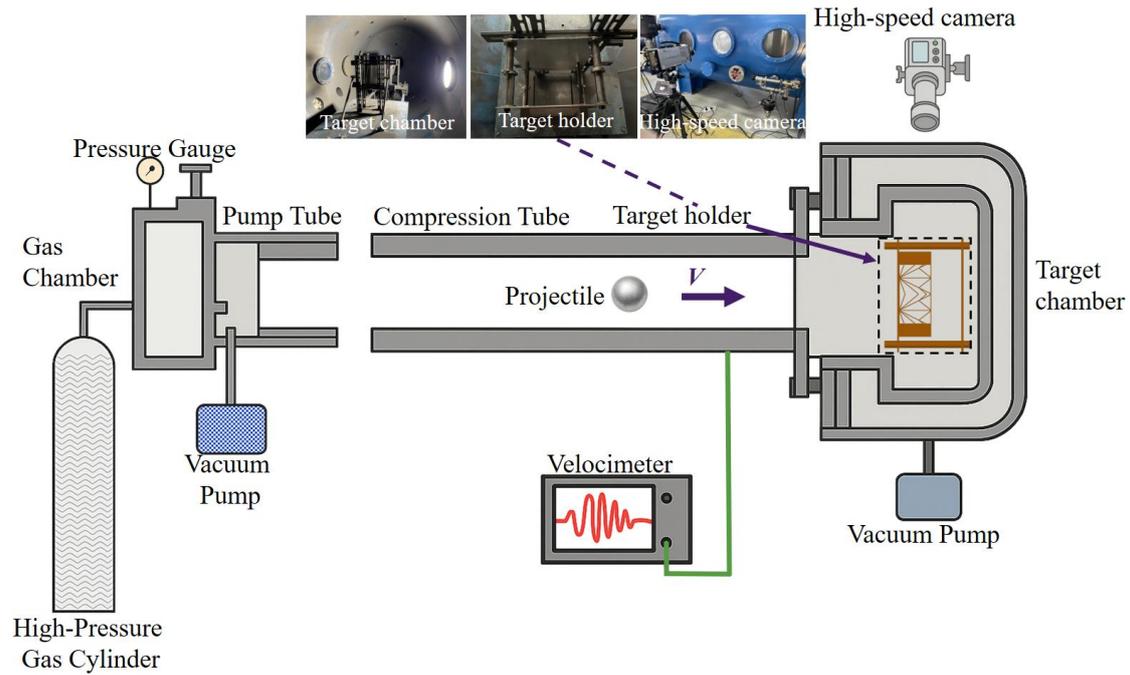

Figure 9. Schematic of the experimental setup and principle.

For the lattice shield structure, three hypervelocity impact experiments were conducted at different incident angles, namely α = 0° (normal incidence), 30°, and 60°. The incident velocity V ranged from 6.0 to 6.5 km/s, and aluminum-alloy projectiles with a uniform diameter of Dp = 5 mm were used. Due to the thin-walled nature of the lattice shield, external observation of internal damage was challenging, and destructive sectioning would have risked additional structural damage. Therefore, to validate the protective performance of this structure, an industrial X-ray computed tomography (CT) system was employed to scan the post-impact lattice shield. The source voltage and current were 220 kV and 210 μA, respectively, and the spatial resolution was 0.148 mm in the x, y, and z directions. The reconstructed volume contained a total of 3,490,226,700 voxels. Post-scan image analysis was performed using the commercial software Avizo to process the tomographic images and reconstruct the three-dimensional structure. SolidWorks three-dimensional geometric analysis software was employed to assess the failure and penetration conditions of the reconstructed structure and to measure the damage dimensions. The failure criterion was defined as material spalling or distinct perforation on the rear surface of the lattice back wall. The critical (near-failure) state was characterized by pronounced delamination and bulging of the back wall, accompanied by a single non-penetrating crack. The detailed test conditions and results are summarized in Table 1.

Table 1 Experimental details and arrangements.

| No. | $α(°)$ | $D_p$ (mm) | $V$ (km/s) | Results |
| --- | --- | --- | --- | --- |
| 1 | 0 | 5.02 | 6.23 | Pass |
| 2 | 30 | 5.02 | 6.33 | Pass |
| 3 | 60 | 5.00 | 6.22 | Pass |

## 3.2 Experimental Results and Discussion

Figure 10 illustrates the damage sustained by the bumper, the rear wall, and the interior of the lattice shield under projectile impacts at incidence angles of 0°, 30°, and 60°. This figure clearly demonstrates the protective performance of the lattice shield and the influence of the impact angle on this performance. When the projectile strikes the lattice shield at 0° (normal incidence) with a velocity of 6.23 km/s, as shown in Figure 10(a), a nearly circular perforation with a diameter of 8.21 mm can be observed on the bumper. The rear wall shows no visible damage. Images from inside the lattice shield indicate that the debris cloud did not reach the rear wall. This suggests that projectiles entering from above collided with the bumper, generating a debris cloud that had already dissipated most of its kinetic energy by the time it reached the fourth and fifth inclined plates, thereby preventing penetration of these layers. As shown in Figure 10(b), when the projectile strikes at an incidence angle of 30° with a velocity of 6.33 km/s, the perforation in the bumper is elliptical, with a major axis of 9.57 mm and a minor axis of 8.23 mm. This behavior is a typical feature of projectile impacts on a target plate at oblique incidence. An intact protrusion with a height of 1.85 mm appears on the rear wall. Images from inside the lattice also reveal an impact crater at the corresponding location on the inner surface of the rear wall. The debris cloud penetrates all inclined plates but is arrested by the rear wall. This trajectory closely matches the theoretical trajectory of the debris cloud within the lattice derived in Section 2, thereby validating the conclusion that an incidence angle of 30° is critical for lattice shielding. Previous studies on Whipple structures have also indicated that, within certain ranges of impact velocity and angle (in particular, at velocities of 5–7 km/s and incidence angles of 30°-40°), oblique impacts cause greater damage to the rear wall than normal impacts at equivalent velocities.

Figure 10(c) illustrates the damage produced by an impact at an incidence angle of 60° and a velocity of 6.21 km/s. At this angle, ricochet occurs when the projectile strikes the bumper, and the ricochet debris cloud creates an impact crater that is visible on the observation plate. The bumper also exhibits an elliptical penetration hole with a major axis of 14.88 mm and a minor axis of 8.86 mm. A comparison with the other two impact-angle cases reveals that larger impact angles result in larger perforation areas. Small elevations on the surface caused by debris impacts are also observed adjacent to the elliptical hole. This is a critical observation, indicating that the inclination of the first inclined plate can deflect a portion of the debris toward directions that are close to the incident direction. This finding is fully consistent with the theoretical derivation in Section 2, as demonstrated by the theoretical debris-cloud trajectory shown in Figure 6(c). Photographs of the rear wall reveal that, because of the large incident angle, the debris cloud was ejected diagonally without passing through the rear wall, leaving no visible damage marks on it.

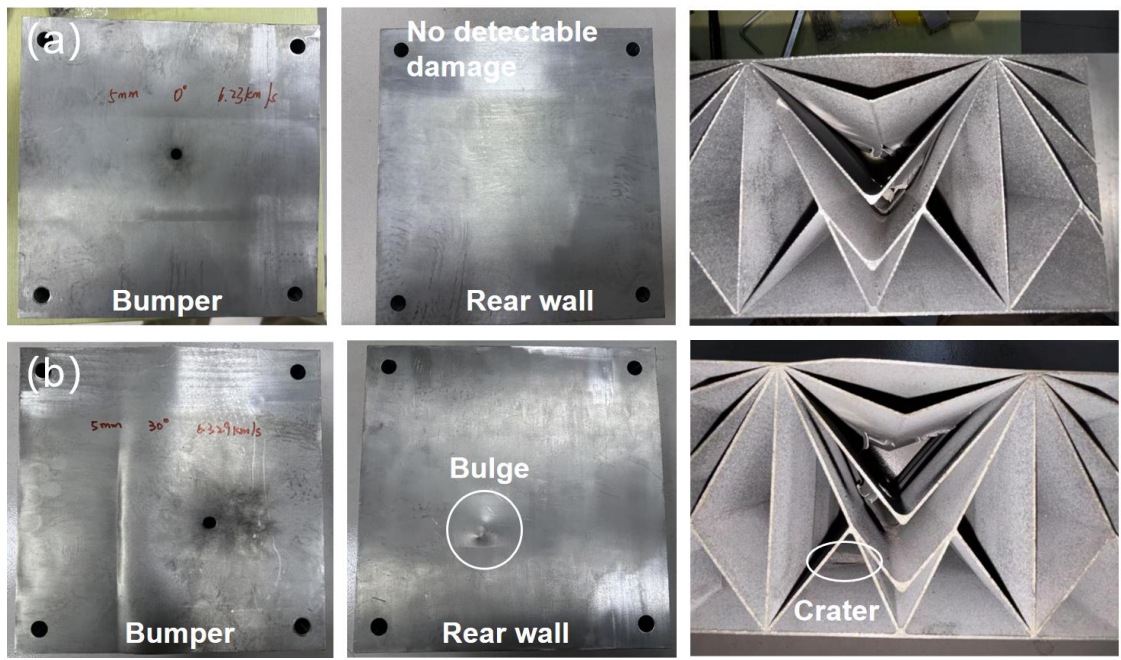

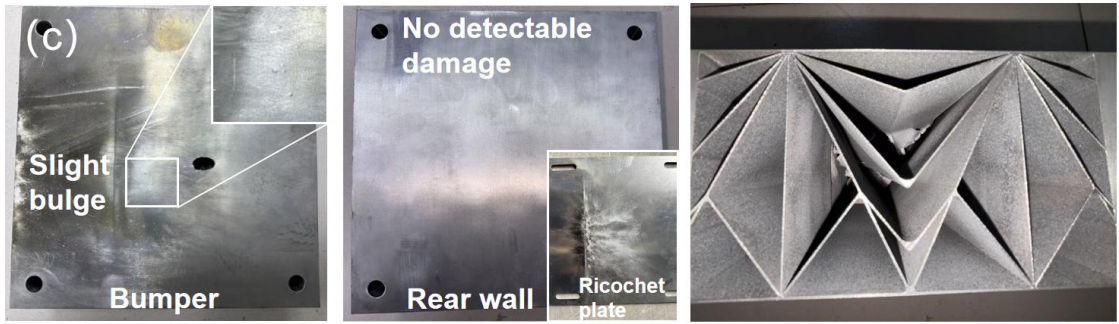

Figure 10. Hypervelocity impact damage morphologies for three incidence angles.

It is difficult to accurately assess the damage state and debris-cloud trajectory from the internal photograph of the lattice structure shown in Figure 10. Segmenting the structure using wire-cutting technology would risk altering the original damage morphology of the thin-walled structure. Therefore, the lattice shield after impact was scanned using X-ray computed tomography (CT), and its three-dimensional geometric model was reconstructed. Details of the scanning and modeling procedures are provided in Section 2. The reconstructed three-dimensional model of the lattice shield after the 30° projectile impact is shown in Figure 11.

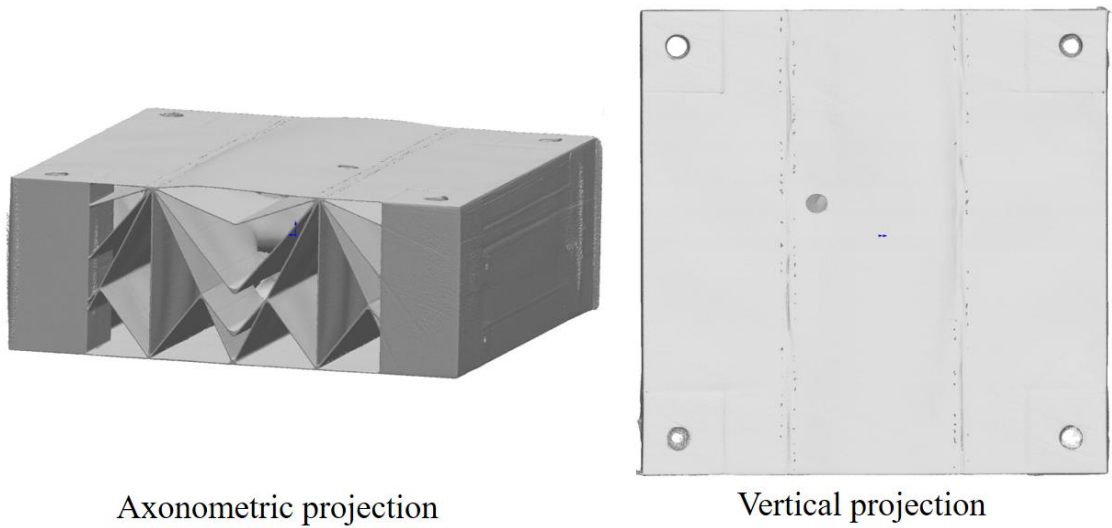

Axonometric projection　　　　　Vertical projection

Figure 11. 3D reconstruction of the lattice shield after impact using industrial micro-CT (30° impact specimen).

The reconstructed three-dimensional model facilitates clear visualization of the shield geometry. The damage dimensions and debris-cloud trajectories can be measured with high precision using the three-dimensional geometric-analysis software SolidWorks and Avizo. The debris-cloud trajectory is evaluated on the basis of the angle at which debris pass through each layer of the shield. The debris transmission angle is measured using the method proposed by William P. Schonberg, as illustrated in Figure 12. In this method, the center of mass and center of momentum of the debris cloud are approximated by the geometric center of the crater formed after the debris cloud collides with the protective plate. The line connecting the geometric center of this crater to the geometric center of the crater on the preceding protective plate defines the debris-cloud trajectory. The angle between this trajectory line and the normal to the protective plate is then measured to quantitatively characterize the debris-cloud trajectory. In Figure 12, $\alpha$ represents the angle of incidence, whereas $\theta_1$ and $\theta_2$ denote the angles between the two resulting debris clouds generated by the projectile impact and the normal to the subsequent protective plate.

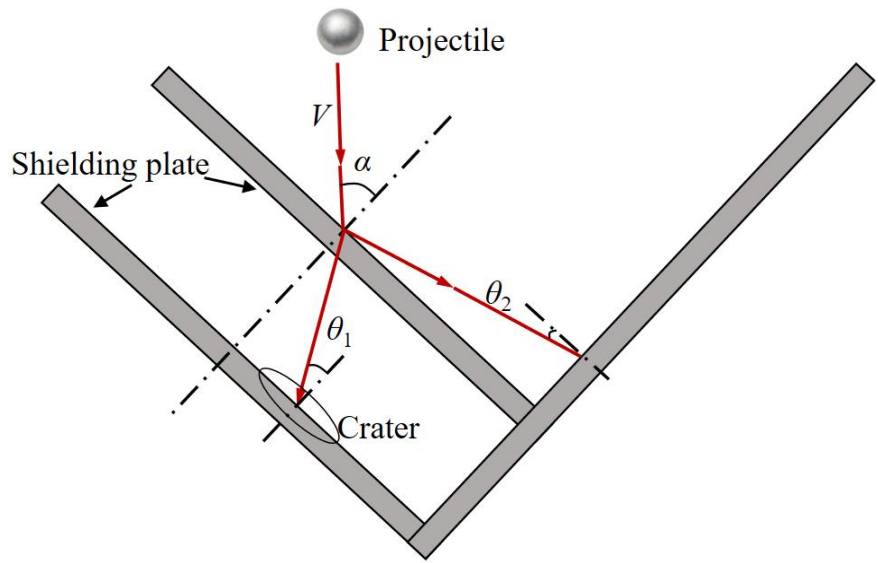

Figure 12. Schematic of the method used to measure debris-cloud trajectory angles.

Using three-dimensional geometric analysis software, the trajectory of the debris cloud during the hypervelocity impact test was measured according to the above method and compared with the theoretical trajectory obtained in Section 2, as shown in Figure 13. The yellow arrows represent experimentally measured data, while the red arrows indicate theoretically derived results. Since Figure 6 already displayed the theoretical debris cloud trajectory angle, only the experimentally measured trajectory angle is labeled in Figure 13. Figure 13 indicates that the theoretical trajectories align well with the experimental trajectories across different impact angle conditions.

Figure 13(a) depicts the trajectory of the debris cloud generated by a 0° impact of the projectile on the shield. The experimentally measured trajectory of the debris cloud agrees closely with the theoretical trajectory, indicating that the theoretical derivation is highly accurate. The measured incidence angle of the debris cloud on the first inclined plate is 23°, whereas the theoretical value is 20°. Considering measurement uncertainty, a difference of 3° is acceptable. As the debris cloud penetrates the first layer and then impacts the second and third inclined plates, deviations from the theoretical trajectory begin to appear. The angle between the normal to the third inclined plate and the experimentally obtained debris-cloud trajectory is 47°, compared with the theoretical value of 51°, corresponding to an 8.5% error. When the debris cloud reaches the side plate, the angle between its trajectory and the plate normal is 39°, which deviates substantially from the theoretical value of 65°. In addition to uncertainties associated with the experimental instrumentation and human operation, three further factors contribute to this discrepancy. First, deviations between the experimental measurements and the theoretical values arise from the initial interaction with the first inclined plate. The start and end locations of each segment of the debris-cloud trajectory affect the measured trajectory angle. Each penetration of an inclined plate introduces an incremental error, so that by the time the cloud reaches the side plate, four such errors have accumulated, leading to the largest deviation from the theoretical prediction at this location. Second, this study approximates the center of mass and center of momentum of the debris cloud by the geometric center of the crater. However, in reality, the geometric center does not coincide with the true center of mass of the cloud. In hypervelocity impact experiments, it is difficult to determine the precise direction of the debris cloud in situ using alternative diagnostic methods. Even with high-speed imaging, the apparent direction of the debris cloud in the images cannot fully represent the direction of its momentum. Finally, each penetration of an inclined plate significantly reduces the kinetic energy of the debris cloud, with the normal component of the velocity decreasing more strongly than the tangential component. Consequently, the normal velocity of the debris cloud when it reaches the side plate in the experiment is already very low, an effect that is not accounted for in the theoretical derivation. Although the theoretically derived trajectory of the debris cloud exhibits some deviations from the experimental trajectory, the overall trend shows good agreement, and the theoretical trajectory captures the actual penetration direction of the debris cloud. In particular, the theoretical prediction of the ricochet debris cloud agrees very well with the experiment, with the discrepancy between theory and experiment for the second and third inclined plates being less than 4°.Figure 13(b) illustrates the debris-cloud trajectory for an impact at an incidence angle of 30°. As analyzed in Section 2, the measured collision angle between the debris cloud and the first inclined plate is 52°, whereas the theoretical value is 50°. The debris then separate into ricochet debris and transmitted debris. The transmitted debris cloud subsequently collides with the following inclined plates at smaller angles, thereby avoiding ricochet, and penetrates multiple inclined plates with an almost unchanged direction before reaching the rear wall, where it causes a certain degree of non-spalling protrusion. At this point, the trajectory line forms an angle of 32° with the normal to the rear wall, having penetrated five layers. The cumulative relative error between the theoretical and measured trajectory angles reaches 28.5%, although the ricochet debris-cloud path remains very close to the theoretical prediction.The schematic of the debris-cloud trajectory for an impact at an incidence angle of 60° is shown in Figure 13(c). The theoretical trajectory of the debris cloud agrees well with the experimentally measured trajectory, with an average angular error of 10.7% across all trajectories. The initial segments of the trajectories are nearly identical in both the theoretical prediction and the experimental measurement. The relative error between the theoretical and experimental trajectory angles for the debris cloud transmitted through the first inclined plate is 32.6%. This discrepancy arises because the low velocity of the debris transmitted by the projectile at 60° incidence makes accurate theoretical prediction of their trajectory more challenging.

The experimental results for the three impact conditions shown in Figure 13 validate the protective performance of the lattice shield. For 5-mm projectiles with velocities in the range 6.2–6.4 km/s and incidence angles of 0°, 30°, and 60°, the shield exhibited no failure. The lattice shield designed in this study effectively disperses the kinetic energy of the debris cloud, significantly reducing its normal kinetic-energy component and deflecting the cloud toward the sides of the shield. When the debris cloud strikes at 0°, it is noticeably deflected to both sides, whereas at incidence angles of 30° and 60° the cloud is debrised into multiple parts by deflection from the inclined plates, thereby substantially reducing damage to the rear wall. Notably, no impact traces are observed on the rear wall for 0° incidence.

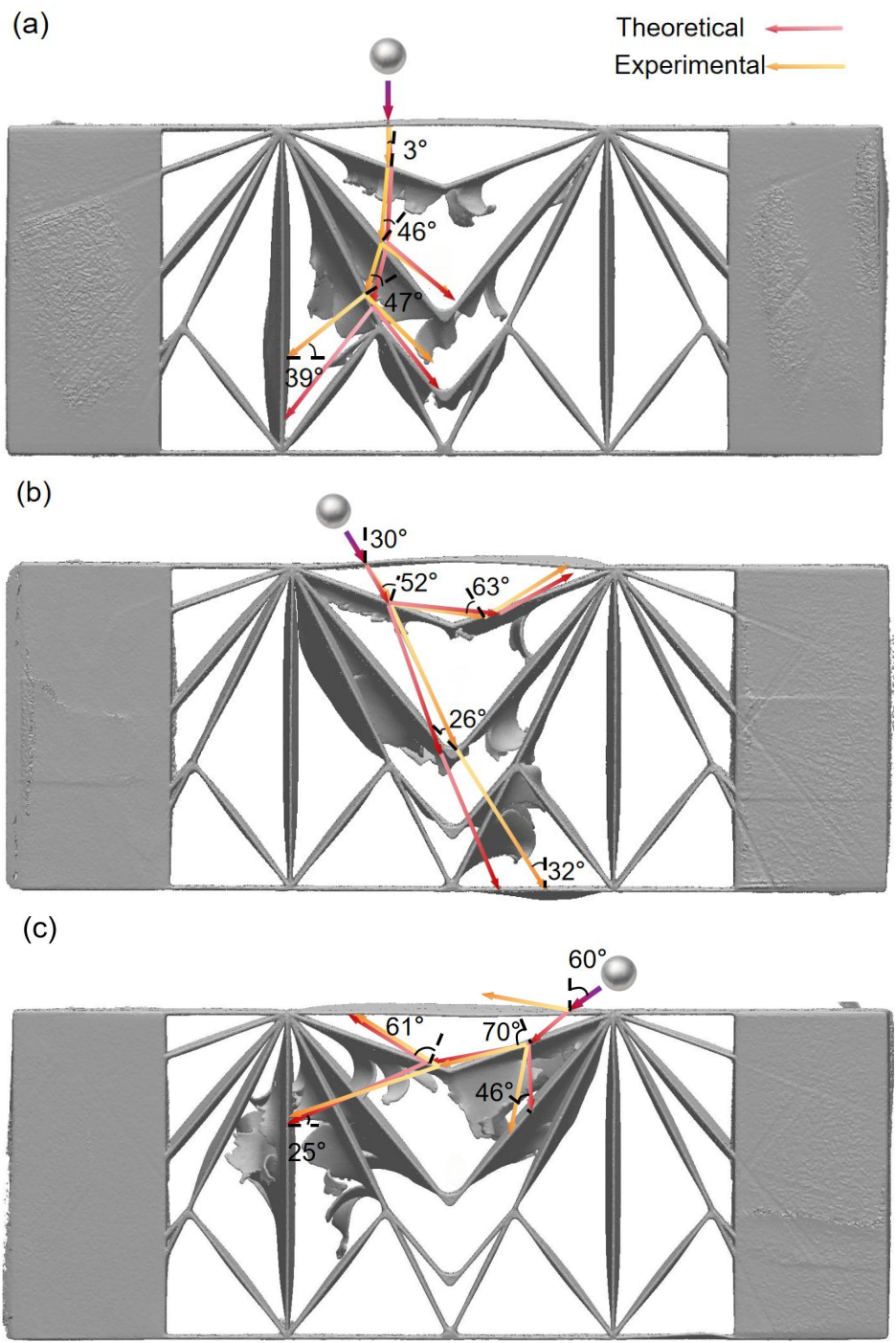

Figure 13. Comparison between theoretical and measured debris-cloud trajectories: (a) 0° impact on the lattice shield; (b) 30° impact on the lattice shield; (c) 60° impact on the lattice shield.

The experimental results also reveal distinct physical mechanisms as the debris cloud penetrates each layer of the lattice shield. Penetration of the bumper produces clean-edged shear holes, whereas penetration of subsequent protective plates causes pronounced flanging at the perforation sites, where the hole-wall material is bent backward and outward. The degree and length of flanging increase with the penetration sequence of the protective plates. This behavior is observed for all three impact angles in the hypervelocity impact experiments. This behavior arises because, under hypervelocity impact conditions of 6.2–6.4 km/s, the impact pressure reaches approximately 80.1 GPa and the strain rate is on the order of $10^6$ s$^{-1}$. The interaction time between the projectile and the target plate is extremely short, rendering material strength properties such as yield strength and shear strength effectively negligible. This impact-driven dynamic effect causes the target plate to behave in a manner similar to that of a fluid. Adiabatic shear occurs at the point of projectile impact, causing severe deformation or even melting of the material surrounding the aperture. The role of shear strength is diminished because the material does not have sufficient time to undergo plastic bending. Instead, it fractures locally under the intense shock wave and high strain rate generated by the impact, resulting in a shear hole that is larger than the projectile diameter. After penetrating the bumper, the projectile and most of the local target-plate material debris, melt, and mix into a high-velocity debris cloud that continues to move rearward. The subsequent inclined plates encounter the debris cloud at specific angles, causing oblique penetration with distinct tangential components and pronounced asymmetry. debris first strike one side of the inclined plate, forcing preferential deformation and bending of that side's material. During impact, this process flips the hole-edge material toward the projectile's trajectory. Holes formed by penetration of the inclined plates often exhibit irregular elliptical or enlarged shapes, with extensive plastic outward deformation along the edges forming a petal-like pattern. This flanging indicates tensile failure of the material. Subsequently, the velocity of the debris cloud decreases from layer to layer. After colliding with the bumper, the projectile is debrised, and a significant portion of its kinetic energy is dissipated. Each subsequent collision with a layer of armor plate results in further loss of kinetic energy. As the number of layers increases, the average velocity and mass of the debris gradually decrease. This implies a reduced peak impact pressure and a prolonged impact duration, allowing the target plate more time to undergo plastic deformation rather than instantaneous shearing. Consequently, the downstream armor plates are more prone to plastic deformation. The further back the armor plate is located, the lower the velocity of the debris cloud it encounters and the greater the spatial extent of the cloud. This reduces the impact pressure per unit area on the armor plate, leading to a higher proportion of plastic deformation and thus more pronounced flanging.

The movement of the debris cloud within the lattice shield is shown in Figure 14. Severe flash phenomena occur during the hypervelocity impact event, primarily caused by localized plasma formation in the aluminum alloy due to the instantaneous high temperatures generated during impact. Given the narrow spacing between layers in the lattice shield, the movement of the debris cloud can only be approximately inferred from the locations where flashes appear. In the 0° projectile impact test, penetration of the third inclined-plate layer is clearly visible at 53.6 μs, indicating that the debris-cloud trajectory entered the side region of the shield along a nearly straight path. High-speed imaging from the 30° and 60° projectile tests shows the debris cloud splitting into ricocheting and penetrating components. At 6.7 μs in the 60° impact test, the first ricochet occurs when the debris cloud strikes the bumper. Ultimately, after two ricochets and multiple penetrations, the debris cloud is dispersed into three distinct parts. The trajectory inferred from the flash positions agrees closely with both the trajectory measured from the crater locations and the theoretical trajectory in Figure 13, providing mutual corroboration.

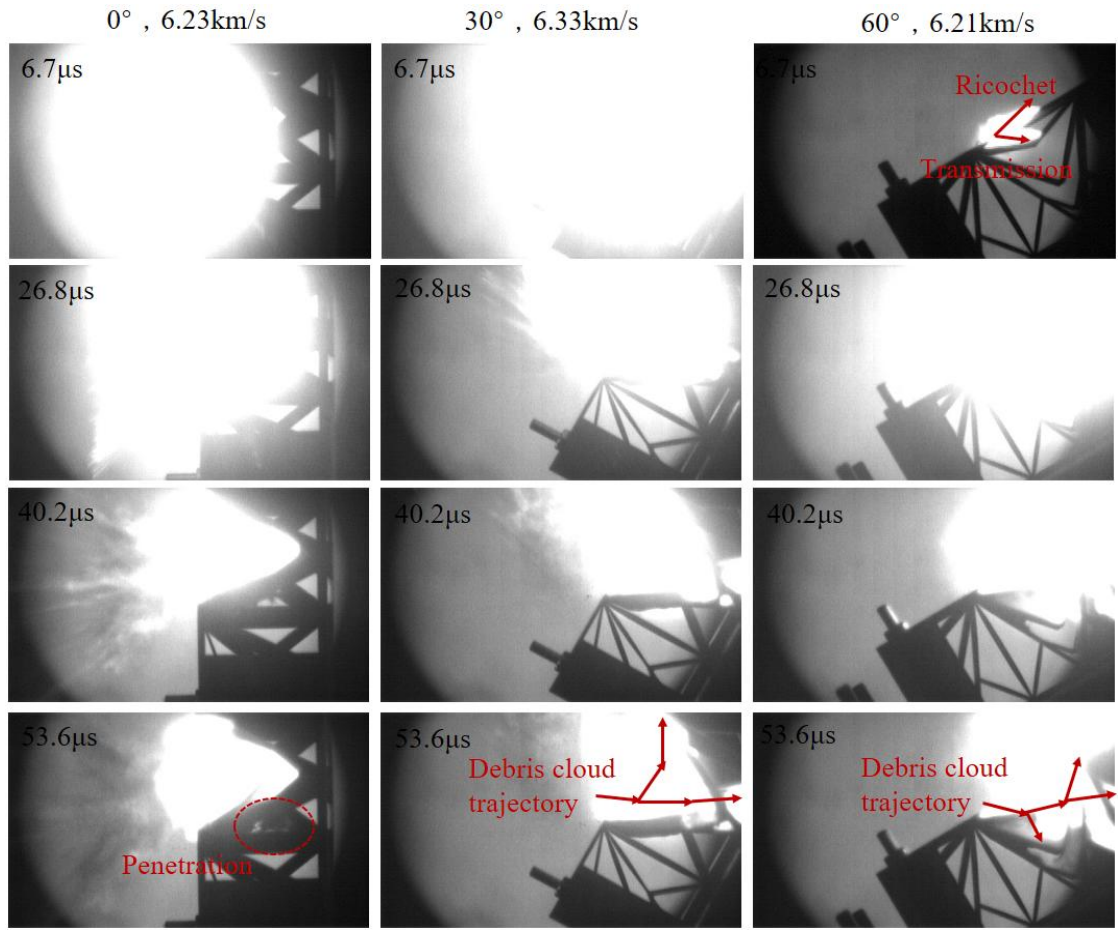

Figure 14. High-speed photographs of debris-cloud trajectories in the lattice shield.

## 4 Numerical Simulation Methods

The small interlayer spacing of the lattice shield, coupled with the generation of numerous flashes during the experiments, makes it challenging to observe in situ the energy-dissipation processes associated with internal structural deformation and debris-cloud evolution. Therefore, smoothed particle hydrodynamics (SPH) numerical simulations were employed to further investigate the hypervelocity impact protection characteristics of the lattice shield.

### 4.1 Numerical simulation model

In this study, SPH simulations were conducted using ANSYS LS-DYNA software to overcome challenges posed by excessive deformation of Lagrangian meshes during severe plastic deformation in hypervelocity impacts. To reduce computational time, the lattice shield was modeled using a single unit cell with a tetrahedral mesh having a characteristic element size of 0.7 mm. This solid mesh was converted to SPH particles using the solid-center method, in which one SPH particle is generated at each mesh center. The projectile was modeled directly in LS-DYNA as a spherical SPH body, with 50 particles along each of the x, y, and z directions. The complete SPH impact model is shown in Figure 4, where red represents the projectile and green denotes the lattice shield. Three hypervelocity impact models were established, corresponding to the experimental impact velocity and angle conditions: 0°, 6.23 km/s; 30°, 6.33 km/s; and 60°, 6.22 km/s. Numerical simulation results were sampled at 1.5 μs intervals to comprehensively capture the trajectory and energy-dissipation process of the debris cloud within the lattice shield.

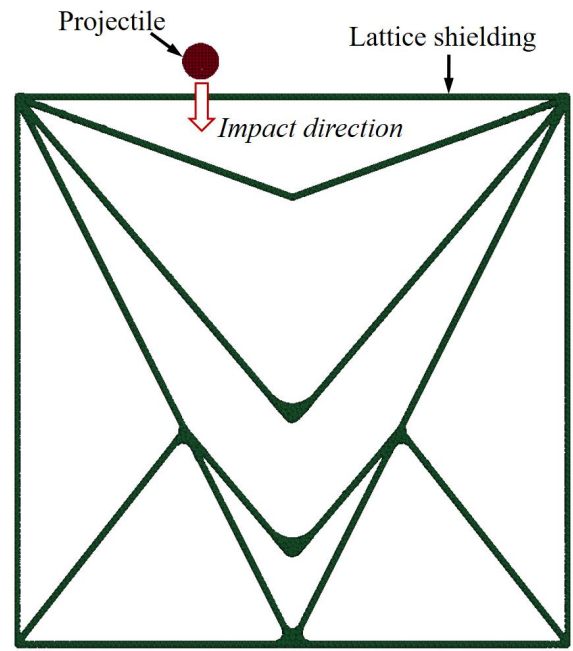

Figure 15. SPH numerical model of the lattice structure.

The projectile material was 6061 aluminum alloy, whereas AlSi10Mg served as the target material. For hypervelocity impact conditions, the selected material model must be suitable for high strain rates. Therefore, the Johnson–Cook (JC) constitutive model was adopted, with parameters derived from tensile tests (as shown in Figure 16) and Hopkinson bar tests performed on SLM-produced AlSi10Mg specimens. The Mie–Grüneisen equation of state (EOS) was employed to calculate the pressures generated during hypervelocity compression. All material parameters used in the simulations are summarized in Table 2. Due to the lack of existing literature on EOS parameters for AlSi10Mg, parameters from aluminum alloys with compositions and properties similar to those of AlSi10Mg were adopted.

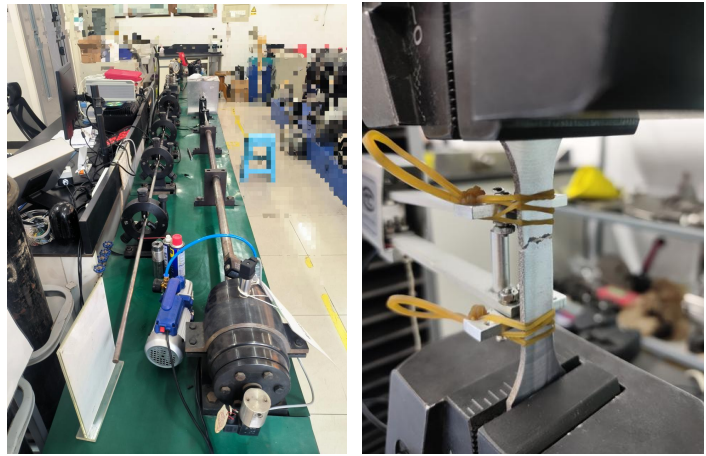

Figure 16. Experiments for calibration of the constitutive model parameters.

Table 2. Material parameters used in the SPH model.

|  | $\rho_0$(g/cm$^3$) | $E$(GPa) | $A$(MPa) | $B$(MPa) | $n$ | $c$ | $C_0$ | $s$ |
|---|---|---|---|---|---|---|---|---|
| SLM AlSi10Mg | 2.67 | 67.1 | 241 | 121 | 0.492 | 0.00943 | 5.32 | 1.37 |
| 6061 aluminum alloy | 2.7 | 71.0 | 324 | 134 | 0.34 | 0.015 | 5.230 | 1.41 |

## 4.2 Numerical simulation verification

A comparison between the numerical simulation results and the experimental results is presented in Figure 17. Figures 17(a), (b), and (c) correspond to the 0°, 30°, and 60° impact conditions, respectively. The right images show the final numerical simulation results, where red represents the projectile debris particles and green indicates the lattice shield and its debris particles. The left images show the corresponding experimental results (see also Figure 13). Since penetration of the protective plates was dominated by projectile debris, the final distribution of simulated projectile debris within the shield was extracted and overlaid on the post-test lattice shield to facilitate visual comparison between the simulations and the experiments. The red particles overlaid on the left-hand experimental images in Figure 17 represent the projectile debris cloud.

A comparison of the debris-cloud particle distribution in Figure 17 with the damage locations on the shield in the experimental images reveals excellent agreement between the SPH hypervelocity impact model established in this study and the experimental outcomes. The simulated damage locations on the shield are nearly identical to those observed in the experiments. When the simulated debris-cloud distribution is overlaid on the experimental images, the regions with concentrated debris particles coincide with the penetration and bulging locations observed in the experiments. However, quantitative discrepancies between the numerical simulations and the experiments arise in the penetration characteristics. This discrepancy is most pronounced for shield plates that are penetrated at later stages in the sequence. For the 0° impact case shown in Figure 13(a), the fourth layer of inclined plates and the side plates are penetrated in the simulation but remain intact in the experiments. A similar discrepancy is observed for the second layer of inclined plates in Figure 13(c). Qualitatively, the numerical simulation of the debris-cloud distribution and trajectories is consistent with the experiments. This consistency arises because the transmission direction and diffusion velocity of the debris cloud are governed predominantly by conservation of momentum, conservation of mass, and the geometric constraints imposed by the thin-walled structure. The inclined-plate configuration deflects the debris cloud at the macroscopic scale. In this regime, the mesh-free SPH method avoids mesh distortion and can stably capture the macroscopic motion of the debris. The Mie–Grüneisen equation of state is effective for computing the propagation velocity of the debris cloud generated by the shock wave and the extent of its expansion after pressure release. In addition, the specific constitutive model has minimal influence on the macroscopic propagation trajectory of the debris cloud. By contrast, damage and penetration thresholds are highly sensitive to the material constitutive response and the damage criteria. At experimental impact velocities approaching 6 km/s, the Johnson–Cook model is not sufficient to fully capture the material flow behavior under hypervelocity impact conditions. The Mie–Grüneisen equation of state also exhibits limitations in representing phase transitions and provides a less accurate description of debris and shield melting or vaporization. Furthermore, the choices of neighborhood radius and artificial-viscosity parameters in the SPH algorithm can affect failure detection. These combined factors lead to lower predicted failure thresholds in the numerical simulations. As a result, significant errors tend to occur in shield plates that are penetrated later in the sequence, because cumulative inaccuracies from the earlier layers eventually give rise to substantial systematic deviations in the final layer.

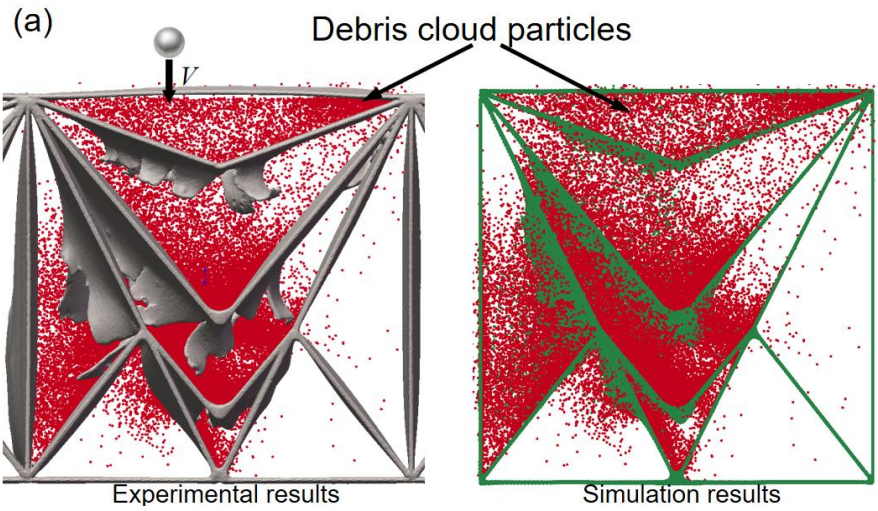

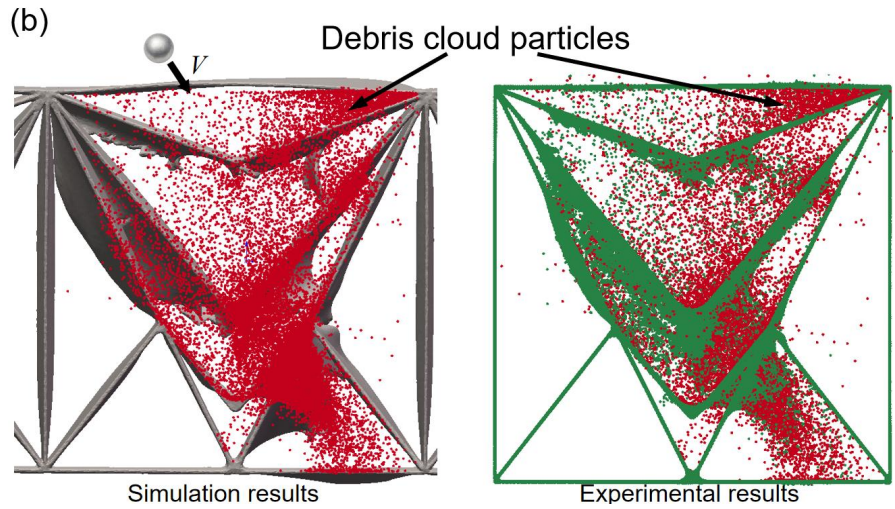

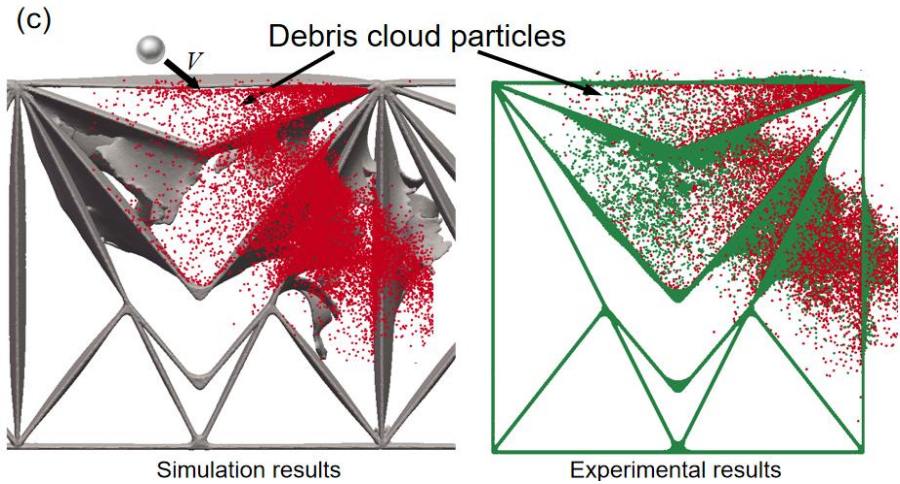

Figure 17. Comparison between SPH numerical simulation and experimental results (left: superposition of the simulated debris cloud on the CT-reconstructed lattice shield; right: SPH simulation result).

## 5 Numerical simulation results and discussion

Section 3 validated the accuracy of the numerical simulation model employed in this study. This section analyzes the energy-dissipation mechanisms of the lattice shield under projectile impact by examining the penetration of the debris cloud through the lattice shield, as revealed by the numerical simulation results.

### 5.1 Debris cloud deflection properties

Figure 18 depicts the penetration process of the debris cloud during an impact at an incidence angle of 0° and a velocity of 6.23 km/s. Figure 19 shows the corresponding velocity-vector field of the debris-cloud particles, where each particle's velocity vector is represented by an arrow. The arrow's size and color indicate the velocity magnitude. After 33 μs, the velocities of all debris in the cloud are nearly zero, and no further penetration of any protective plates occurs. At 3 μs, the projectile penetrates the buffer plate. The enlarged view shows the onset of debrisation, although a large-mass core debris retains most of the momentum. At 6 μs, the central large debris and the surrounding cloud penetrate the first inclined-plate layer. The central debris undergoes secondary debrisation due to reflection from rarefaction waves and collision with the protective plate. The velocity-vector field in Figure 19 reveals that the impact at 6 μs on the first inclined plate generates ricochet debris. A small number of particles within the ricochet debris cloud exhibit extremely high velocities. Experimental results reported in the literature indicate that a very small fraction of low-mass debris within ricochet clouds can achieve velocities significantly exceeding those of the original debris, thereby corroborating the findings of this numerical simulation.

During penetration of the successive inclined plates, the collision angles of the debris cloud undergo significant changes. Upon impacting the second and third inclined plates, the debris do not penetrate directly but slide along the plate surfaces, acquiring lateral velocity components that cause an overall deflection of the cloud. As analyzed in Sections 2 and 3, the collision angle between the debris cloud and the second inclined plate is approximately 50°. This interaction generates a large number of ricochet debris in addition to the transmitted debris, as confirmed by the simulation snapshot at 9 μs. The ricochet debris generated on the left side of the second inclined plate perforate its right side, while the transmitted debris cloud continues toward the third inclined plate. The velocity-vector field reveals that, after striking the first two inclined plates, the debris cloud's velocity direction has deviated significantly from its initial incidence angle. Its impact on the third inclined plate mirrors the interaction with the second plate, again splitting the cloud into two components. However, the collision angle at the third plate is greater than at the second plate, resulting in ricochet debris of higher mass and velocity. The fourth inclined plate is specifically designed to shield against these ricochet debris generated at the third plate. The velocity-vector field shows that, by 15 μs, the debris cloud velocity has decreased significantly to approximately 2 km/s (green/yellow arrows) upon penetrating the third inclined plate. This observation is consistent with the experimental results and analysis in Section 3: the debris cloud velocity continuously decreases during penetration, causing the failure mode of the protective plates to gradually shift from phase transformation and brittle fracture to plastic deformation. Beginning at 21 μs, the leading edge of the debris cloud transmitted through the third inclined plate deflects at a larger angle toward the side plates, ultimately causing their plastic deformation. The velocity of the residual debris is insufficient to penetrate any protective layer, which manifests as a slow diffusive motion of the cloud.

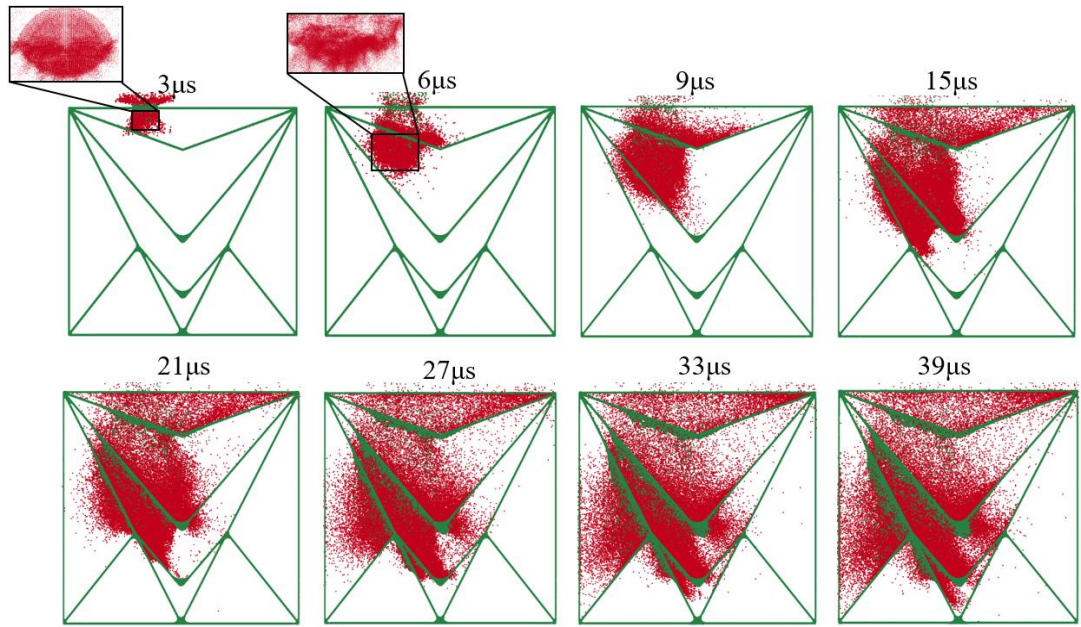

Figure 18. SPH-simulated debris-cloud trajectory for a 0° impact.

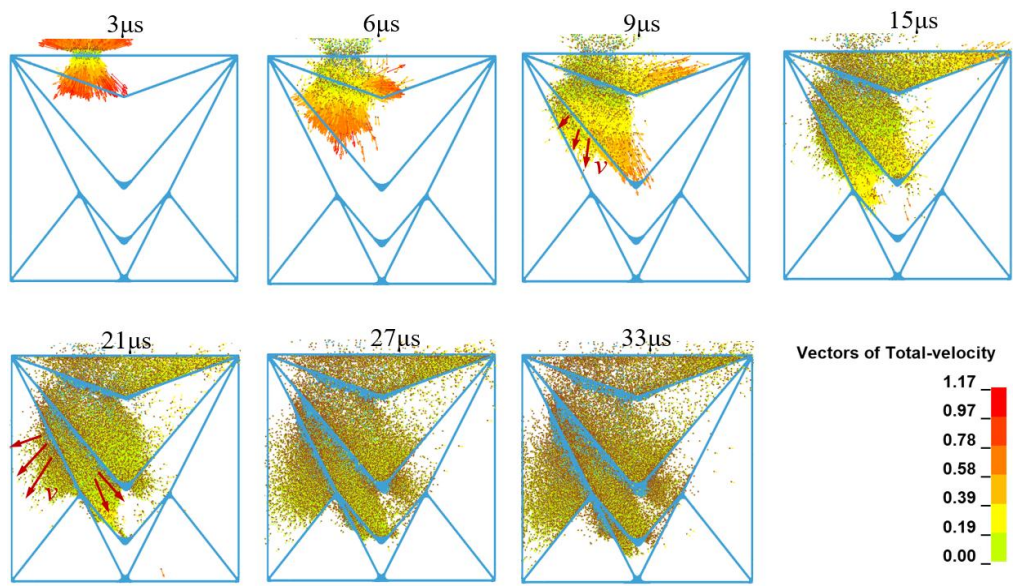

Figure 19. SPH-simulated debris-cloud velocity vector field for a 0° impact.

Figure 20 shows the trajectory and velocity-vector field of the debris cloud from the numerical simulation for an impact at an incidence angle of 30° and a velocity of 6.33 km/s. Since oblique impacts at certain angles cause lateral debris within the cloud to concentrate more strongly, collision angles between 30° and 40° represent a critical risk range for shields. At these angles, Whipple-type shields experience more severe damage from projectile impacts. The magnified view at 3 μs reveals that the projectile trajectory intersects the target plane along an elliptical path due to the misalignment between the projectile axis and the plate normal. Consequently, the penetration hole in the buffer plate observed in the experiment shown in Figure 10(b) is elliptical. At 9 μs, the central large debris generated by the buffer-plate collision ricochets from the left side of the first inclined plate at high velocity, driven primarily by tangential momentum. Because the effective impact angle is relatively small, transmitted debris clouds still dominate. At 21 μs, the debris ricochets from the second inclined-plate layer at a large angle and penetrates downward at high velocity. The ricochet flux and the associated velocity components increase accordingly. However, the subsequent collision with the protective plate occurs at a small angle, resulting in negligible additional ricochet and trajectory deflection. After energy dissipation through multiple protective layers, the kinetic energy of the debris cloud is significantly depleted. Upon striking the rear wall, the debris cloud causes only plastic deformation without penetration. At 39 μs, the velocity-vector arrows of the debris cloud have become smaller and appear green, indicating that the velocity is approaching zero.

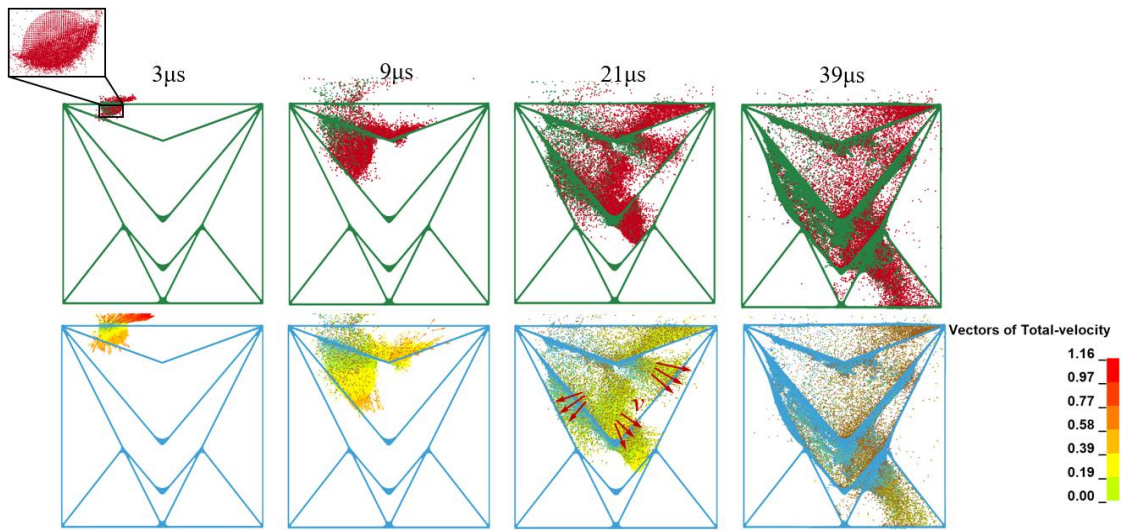

Figure 20. SPH-simulated debris-cloud trajectory and velocity vector field for a 30° impact.

Figure 21 shows the trajectory of the debris cloud during an impact at an incidence angle of 60° and a velocity of 6.23 km/s. At 3 μs, the projectile collides with the buffer plate. The 60° incidence angle exceeds the critical angle of 45° for ricochet generation, resulting in a large number of ricocheting debris. The velocity-vector field indicates that the velocities of the ricocheting debris are significantly greater than those of the transmitted debris. After the debris reach the left side of the first inclined plate, the analysis in Section 2.3 indicates a collision angle of 52°, which triggers a second ricochet. At 9 μs, the ricocheted debris collide with the right side of the first inclined plate, causing a third ricochet. The images at 9 μs and 10.5 μs show the debris cloud penetrating at high velocity toward the upper-left, explaining the crater formation adjacent to the buffer plate in Figure 10. debris that penetrate the first inclined plate have already decelerated substantially. After multiple subsequent collisions with the protective plates, their velocity magnitudes decrease further and their trajectories are slightly deflected, so that they can no longer reach the rear wall.

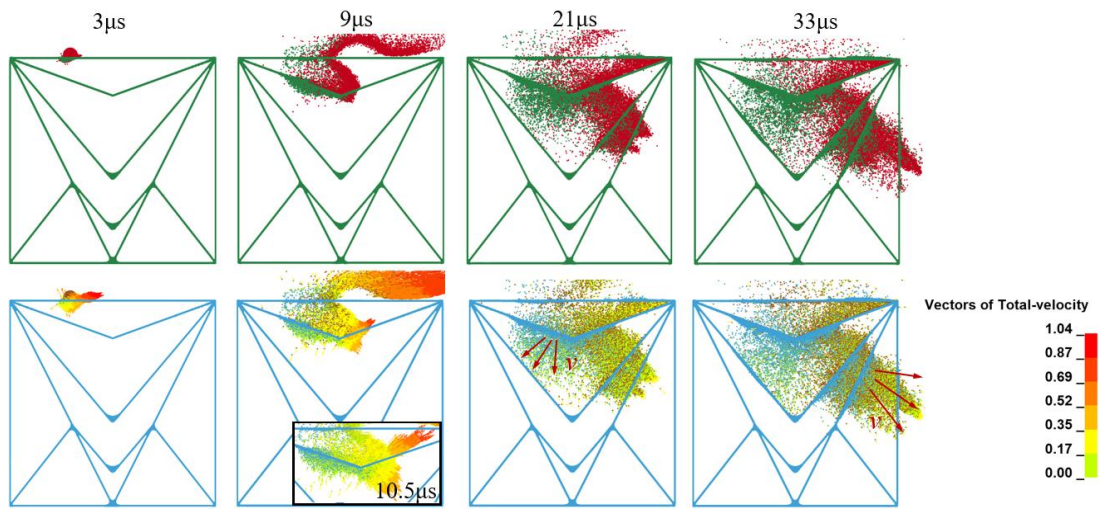

Figure 21. SPH-simulated debris-cloud trajectory and velocity vector field for a 60° impact.

As shown in Figures 19–21, when debris strike nearly normal to the shield, interactions with each inclined plate cause further debrisation and dispersion. The velocity vectors diverge significantly, greatly reducing the concentration and penetration capability of the debris cloud. Owing to the gradient arrangement of the inclined-plate angles, the normal momentum of the debris cloud is progressively dissipated through phase transformations, penetration, and plastic energy dissipation during each interaction. At the same time, a certain amount of tangential momentum is imparted, causing a systematic lateral deflection of the overall trajectory. For 30° and 60° impacts, the debris cloud reaches or approaches the critical angle for ricochet at the first inclined plate, forming a distinct ricochet debris cloud with substantial momentum redirected diagonally upward. In addition, the oblique incidence causes the normal component of the projectile or debris velocity to be inherently low, and it decays further during subsequent plate collisions. Ultimately, the normal impulse is insufficient to cause penetrative failure of the rear wall, resulting only in localized plastic bulging. Overall, the lattice shield effectively reduces the impact kinetic energy of the debris cloud through the coupled action of dispersion and deflection–induced energy dissipation provided by its gradient inclined plates.

5.2 Debris cloud kinetic properties

Section 4.1 qualitatively analyzed the energy-dissipation pathways of the debris cloud along its propagation path within the lattice shield. Figure 22 shows the kinetic-energy evolution of the system under a 0° normal impact condition, quantitatively revealing the redistribution of system energy after impact. As depicted in Figure 22, the initial total system energy is equal to the kinetic energy of the projectile (3.4 kJ). At 1.5 μs, the projectile collides with the buffer plate, generating a shock pressure of approximately 80 GPa, which significantly exceeds the dynamic phase-transformation threshold of aluminum alloys. The debrisation and phase transformation of both the projectile and the target plate effectively dissipate the impact energy. The remaining energy remains stored as kinetic energy within the transmitted debris and the shield, causing the shield's kinetic energy to rise from zero. However, the projectile retains substantial kinetic energy. At 4.5 μs, projectile debris collide with the first inclined-plate layer. The central large debris fractures, dissipating substantial energy. In this stage, the hypervelocity impact causes debrisation and phase transformation in both the projectile and the first inclined-plate layer. Kinetic energy is efficiently converted into internal energy and into the kinetic energy of the debris and the shield, causing the shield's kinetic energy to rise to its peak value. After 4.5 μs, because of energy dissipation in the previous two impacts, the projectile's kinetic energy is no longer transferred efficiently to the lattice shield. Consequently, the lattice shield's kinetic energy begins to decrease, and the rate of decline of the projectile's kinetic energy is reduced. At 10.5 μs, the projectile collides with the second inclined-plate layer. With the projectile kinetic energy significantly reduced, the shield plate primarily undergoes plastic failure, entering a plastic-deformation–dominated energy-absorption stage. Compared to the initial instant, the rate of kinetic energy conversion to other forms markedly decreases during this phase. This is because the debris have been substantially decelerated by prior collisions, subsequently inducing plastic deformation in the target plate at lower velocities. Plastic deformation energy dissipation is a relatively gradual mechanism, with material flow and yield absorbing energy at comparatively low rates.By 27 μs, when the ricochet debris cloud and the transmitted debris cloud collide with the third inclined plate and the side plate, respectively, the total kinetic energy of the system is nearly reduced to zero. Although the lattice structure absorbs most of the impact energy, the residual kinetic energy of the system in the simulation does not completely drop to zero. This residual kinetic energy arises because a minute fraction of debris within the impact cloud may remain in motion at the simulation's conclusion, either not fully intercepted or failing to collide with the structure. The total kinetic energy carried by these debris is negligible, yet it remains undissipated within the finite simulation timeframe. In real-world experiments, the residual kinetic energy of these minute debris is ultimately dissipated as heat through air drag or repeated collisions with the structure. However, in simulations, the kinetic energy of these residual debris persists as a residual value as long as they do not undergo further interactions.

It should also be noted that shock waves generated by the impacts dissipate energy through repeated reflection and transmission within the structure. In lattice structures, each layer of plates progressively attenuates the kinetic energy of the debris cloud. The gradient angle design effectively prevents any single layer from bearing the full impact energy of the debris cloud, thereby enhancing the overall resistance to penetration.

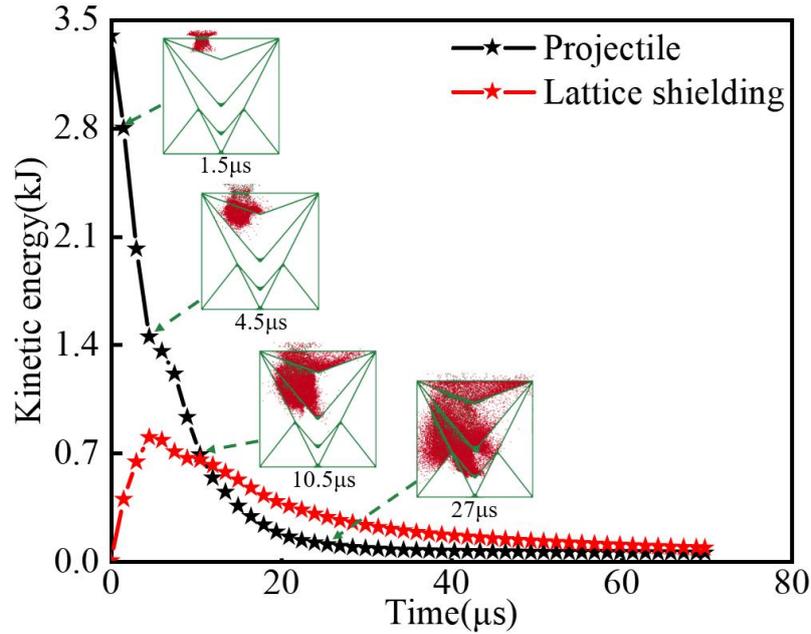

Figure 22. Numerically simulated kinetic-energy histories of the projectile and the lattice structure for a 0° impact.

To clarify how the lattice shield disperses and dissipates the energy of each debris-cloud segment, the debris cloud is divided into four parts according to trajectory: transmitted debris, ricochet debris, ricochet debris 2, and ricochet debris 3. For each of these four segments, a representative debris particle is selected, and the velocity–time trajectories of these particles are plotted, as shown in Figure 23. For a 0° incidence angle, the transmitted debris cloud accounts for the largest mass fraction among the four segments. At t = 0, the projectile velocity upon impact with the shield decreases from A to B. After the first layer of plates fractures, the debris cloud moves downward through the interlayer gap. During this motion, the debris do not contact the surrounding medium or structure, and thus their kinetic energy neither performs work on the structure nor is converted into internal energy. This produces a plateau segment in the curve. Because the selected particle may penetrate the first inclined plate later than the preceding particles, its velocity decreases only slightly upon colliding with the second inclined plate at 3 μs, resulting in a nearly horizontal BC segment. Points C and D correspond to the velocity decay following collisions with the second and third inclined plates, respectively. After slowing to 575 m/s, the debris undergoes a period of essentially free motion, represented by the velocity plateau segment between points D and E, until its velocity decays to nearly zero upon colliding with the side plate at point E. debris 1 is generated when the debris cloud collides with the first inclined plate. Due to the small angle of impact with the first plate, debris 1 represents only a very small mass fraction and does not cause penetrating damage to subsequent protective plates. Point F corresponds to its collision with the left side of the first inclined plate. As established in this work and in the literature[41], a small number of ricochet debris can exceed the projectile's initial velocity. Consequently, a pronounced peak velocity of 8.9 km/s is observed at point G. This quantitatively supports previous assertions that this phenomenon arises from shock waves formed at the collision interface, which reflect as tensile waves at the material boundary and instantaneously accelerate the local material. At point G, where contact occurs with the right side of the first inclined plate, the debris velocity sharply decreases. After ascending to point H and colliding with the buffer plate, the debris subsequently undergoes multiple collisions within the confined space. Ricochet debris cloud 2 is generated when the transmitted debris cloud collides with the second inclined plate at point I, where a smaller jump in the velocity signal indicates that a ricochet has occurred. Points J and K represent the times at which ricochet debris 2 collides with the right side of the second inclined plate and with the third inclined plate, respectively. Ricochet debris cloud 3 is generated when the transmitted debris cloud collides with the third inclined plate. Point L marks the onset of the generation of ricochet debris cloud 3 and corresponds to point D. Subsequently, debris cloud M collides with the fourth inclined plate and then penetrates through the first and second layers before its energy is fully dissipated.

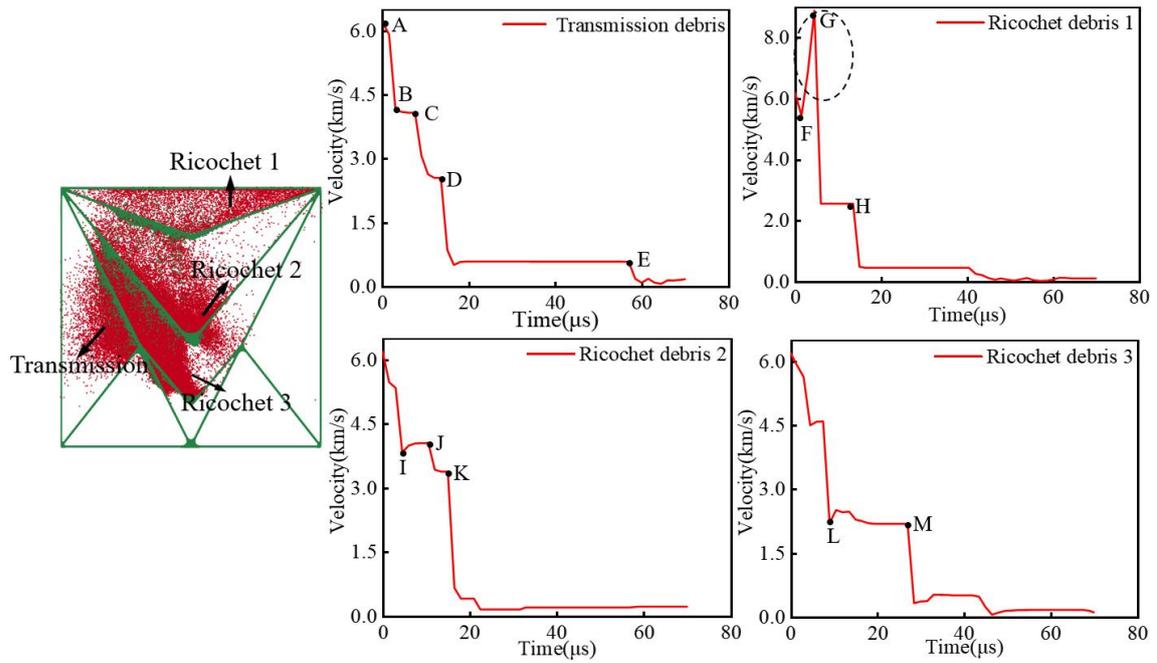

Figure 23. Numerically simulated velocity–time histories of the different debris-cloud components for a 0° impact.

Figure 24 shows the kinetic-energy evolution curve for the 30° impact case. As in the initial segment of the 0° impact kinetic-energy curve, the projectile kinetic energy rapidly decays, while a portion is coupled into the lattice structure, causing the structural kinetic energy to rise sharply from zero. By 13.5 μs, ricochet debris clouds collide with the buffer plate, while debris from the first inclined-plate layer collide with the second inclined-plate layer, generating new ricochet debris. These two types of debris produce a new round of debrisation, collision-induced heating, and plastic-deformation energy dissipation, leading to a rapid decrease in kinetic energy. Correspondingly, the kinetic energy input to the lattice structure increases, leading to a brief rebound in structural kinetic energy. By approximately 34.5 μs, the kinetic energy of the debris is largely dissipated, and the residual velocities are insufficient to cause penetration of the rear wall. Consequently, only a non-penetrating plastic response of the rear wall is observed. Notably, under the 0° impact condition, the projectile kinetic energy is nearly fully dissipated by 27 μs. However, under the 30° impact condition, this dissipation time increases by 25.9%. The underlying mechanism is that, at a moderate slope, the normal component of the impact velocity is lower than that of a normal impact but still sufficient to penetrate the next layer.

Consequently, energy transfer to the structure is delayed, requiring a longer duration for overall dissipation. Compared with the 0° condition, the 30° impact more readily forms a nearly straight penetration path, resulting in a slower rate of energy dissipation. Conversely, under the 0° condition, the gradient multi-layer inclined-plate configuration achieves energy dissipation and ricochet within a short timeframe, leading to a higher energy-decay efficiency.

To clarify the energy-transfer path, the debris cloud is classified according to its motion characteristics, and representative particles are selected to construct velocity–time curves, as shown in Figure 25. Two debris-cloud components are considered: the ricochet debris cloud and the transmitted debris cloud from the first inclined-plate layer. Point A corresponds to the time t = 13.5 μs in Figure 23. The segment between points A and B shows a reduced slope of the velocity–time curve, resulting from the direct interaction of transmitted debris with multiple layers of protective plates during their motion. Point C corresponds to the collision with the rear wall, where the velocity drops to nearly zero, indicating that the particle's energy has been almost completely dissipated and that penetration of the protective plates is unlikely. In the ricochet-debris velocity curve in Figure 24, a distinct jump in the velocity signal appears before point D. Its cause and timing are identical to those for the ricochet debris in Figure 23. By point E (t = 40 μs), the debris particle's velocity has dropped to zero upon impact with the buffer plate. The curves in Figures 25 and 24 corroborate one another, demonstrating that the lattice shield achieves effective protection against the debris cloud through multi-stage energy dissipation by dispersion and deflection even under moderate-slope conditions.

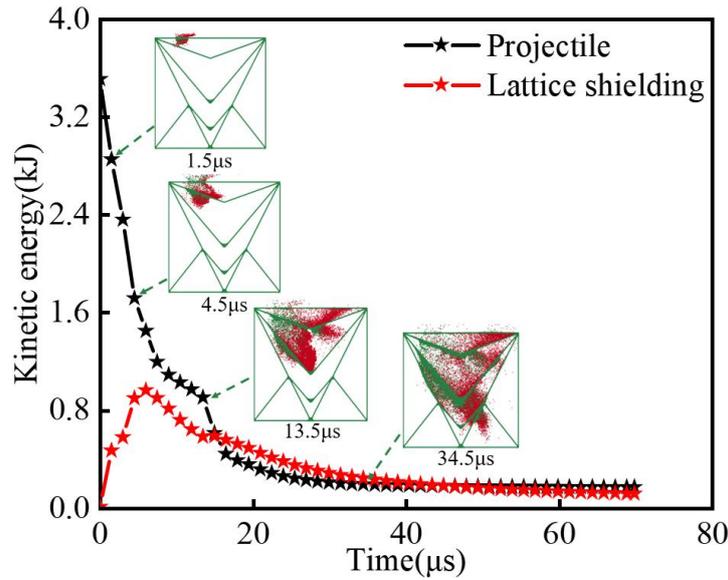

Figure 24. Numerically simulated kinetic-energy histories of the projectile and the lattice structure for a 30° impact.

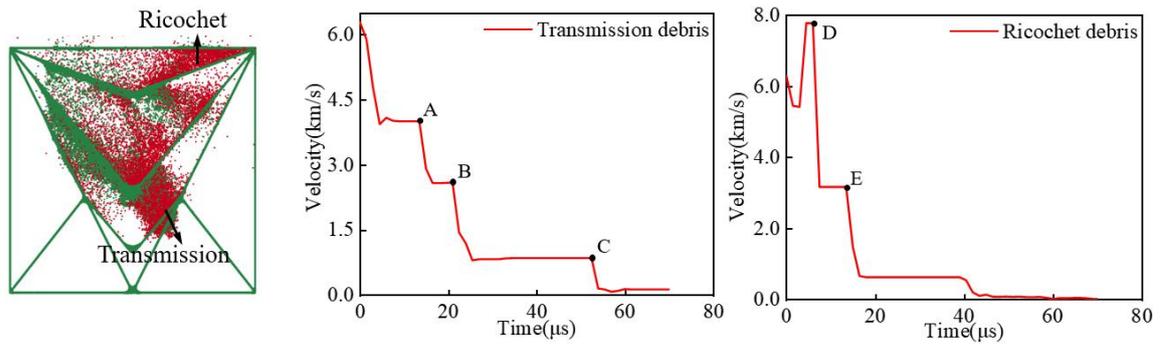

Figure 25. Numerically simulated velocity–time histories of the different debris-cloud components for a 30° impact.

Figure 27 shows the kinetic-energy evolution of the system for the 60° impact case. At 25.5 μs, the transmitted debris cloud penetrates multiple protective plates before colliding with the side plate. The time required for the kinetic energy to reach its minimum value exceeds that for the 0° and 30° impacts. This is because the 60° incidence angle corresponds to a steep oblique impact with an extremely low normal momentum component, under which ricochet-dominated debrisation occurs. Most of the kinetic energy is converted into ricochet energy during the initial impact on the buffer plate (as indicated by the 6 μs debris trajectory), resulting in a stepwise multi-stage decrease after penetration of multiple layers. Simultaneously, the debris ricocheting at the buffer plate and those penetrating the side plate move freely in the numerical simulation because no subsequent protective plates are present. Consequently, by the end of the simulation, the projectile's kinetic energy remains at approximately 0.9 kJ, despite having stayed near its minimum value for an extended period. From an engineering standpoint, this residual energy does not pose a risk of perforating the protective structure.

The particle velocity–time trajectory curve is plotted in Figure 28. The primary debris cloud components originate from ricochet debris and transmitted debris generated upon collision with the first inclined-plate layer. The kinetic-energy dissipation mechanism of the ricochet debris is similar to that at a 30°incidence angle, but their velocity and mass contributions are higher, enabling penetration through the buffer plate. The velocity curve of the transmitted debris cloud drops sharply to 2.4 km/s at point A. After colliding with the right side of the first inclined plate at point B, it exhibits several stepwise decreases, each step corresponding to collisions with the second and third inclined plates and the side plate.

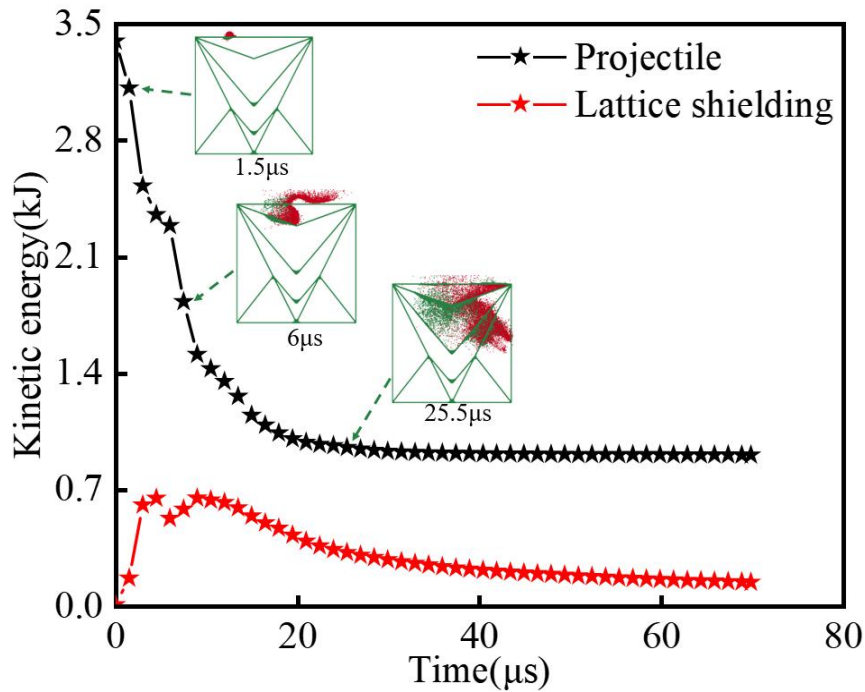

Figure 26. Numerically simulated kinetic-energy histories of the projectile and the lattice structure for a 30° impact.

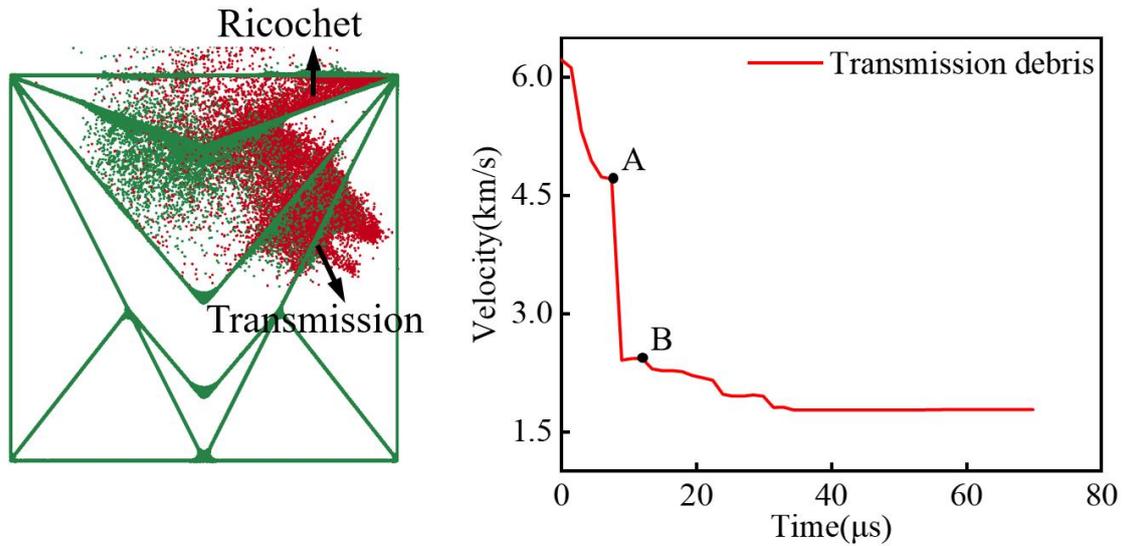

Figure 27. Numerically simulated velocity–time histories of the debris cloud for a 30° impact.

In summary, the lattice structure achieves significant mitigation of hypervelocity impacts through the coupled effects of hierarchical energy dissipation and guidance of debris kinetic energy.

## 5.3 Ballistic limit of lattice shielding

To verify the deflection and kinetic-energy dissipation effects of the lattice shield on projectile trajectories at different impact positions, numerical simulations were conducted for projectiles striking at various locations on the shield. In Figure 28, numerical simulations are presented for a 5 mm projectile traveling at 6.2 km/s impacting at three representative positions located 0 mm, 12.5 mm, and 22 mm from the lattice's central axis. In Figure 28(a), when the projectile strikes along the lattice's central axis, each protective layer disperses the debris cloud laterally. Consequently, the debris cloud exhausts its kinetic energy before penetrating beyond the fourth inclined-plate layer. For the off-axis impact points depicted in Figures 28(b) and 28(c), the lattice shield similarly provides effective protection, preventing the debris cloud from penetrating the rear wall. It should be noted that numerical models typically yield conservative penetration predictions relative to experiments, owing to limitations in the material constitutive models and particle search algorithms. The fact that non-penetration was still predicted under these conditions indicates that the lattice shield provides a high margin of protection over a range of impact locations, suggesting an even greater safety margin in practical engineering applications.

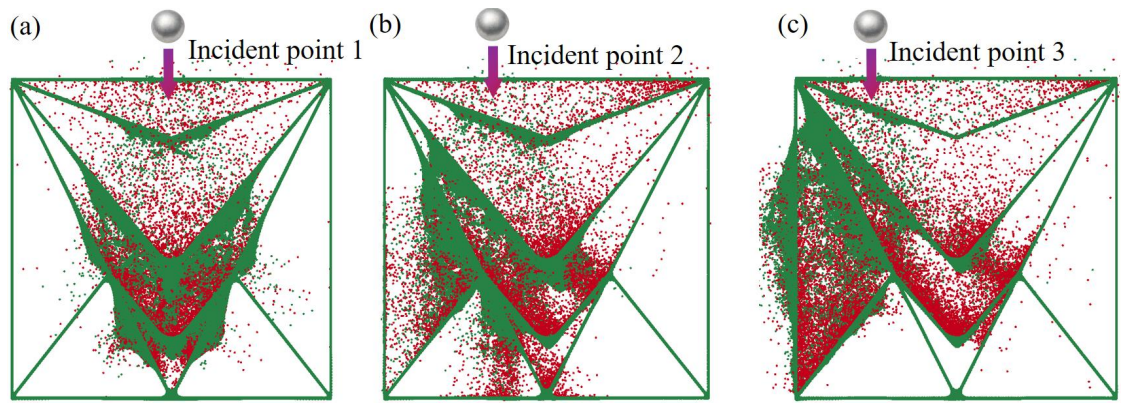

Figure 28. Protective performance of the lattice shield for different impact locations.

The ballistic-limit curve characterizes the relationship between the critical projectile diameter and the impact velocity associated with failure of the protective structure. Points above the curve correspond to failure, whereas those below correspond to no failure. At a given projectile velocity, a larger critical diameter signifies greater protective capability. As shown in Figure 29, despite the limited number of experimental data points, comparison of the ballistic-limit curves for the lattice shield and a traditional 6061 aluminum alloy Whipple structure — using Whipple data reported in the literature — reveals that the trajectory-planning lattice shield exhibits an overall upward shift of its ballistic-limit curve[43]. This demonstrates superior protective performance compared with the Whipple shield, with an average increase of 29.1% in the maximum projectile diameter that can be safely intercepted.

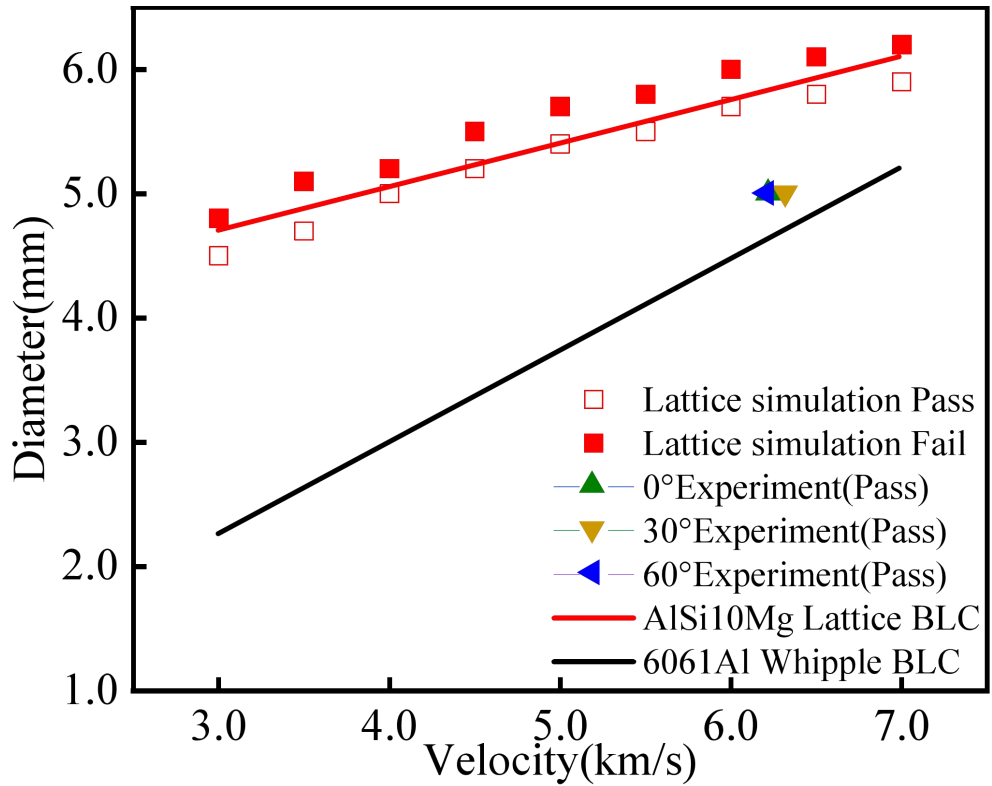

Figure 29. Comparison of ballistic-limit curves for the Whipple shield and the lattice shield[43].

# 6 Conclusion

This study proposes and validates a trajectory-planning lattice protective structure additively manufactured with aluminum alloy. Guided by the design principles of debris-cloud deflection and hierarchical energy dissipation, this structure achieves superior hypervelocity protection performance while significantly reducing overall thickness compared with traditional aluminum alloy Whipple structures. Experimental data from the literature, together with the extensive numerical simulations conducted in this work, demonstrate that the transmission direction of projectiles subjected to hypervelocity oblique impact on target plates is deflected. When the incidence angle exceeds a critical value of approximately 45°, a ricochet debris cloud dominated by tangential momentum emerges, and the fraction of kinetic energy carried by this cloud increases as the incidence angle increases. On this basis, a functional relationship is established between the incidence angle and the transmission and ricochet trajectory angles. This relationship enables customized design of the protective-structure geometry to control the trajectory of the debris cloud and allows derivation of the theoretical debris-cloud trajectory within the lattice. Hypervelocity impact experiments were performed to evaluate the protective capability of the lattice structure. In oblique hypervelocity impact tests at 0°, 30°, and 60° with incidence velocities of 6.2–6.4 km/s, the lattice protective structure did not fail and exhibited outstanding protective performance. Industrial CT scanning of specimens, combined with statistics of perforation locations and damage morphologies, confirmed that this structure significantly deflects the momentum direction of the debris cloud toward the lattice sides while effectively dissipating its kinetic energy. The experimentally observed trajectories are in good agreement with the theoretical predictions.

Based on the SPH numerical model, the protective mechanism of the lattice structure is further elucidated: inclined plates with gradient angles progressively weaken the normal momentum of the debris cloud through continuous debrisation and dispersion together with trajectory deflection, while simultaneously introducing a tangential component. This causes the momentum direction to change continuously and the kinetic energy to dissipate in a stepwise manner. The energy-dissipation process exhibits distinct characteristics at different incidence angles. Compared with the 0° and 60° cases, the 30° impact results in the least effective energy dissipation and the lowest efficiency because fewer ricochets of the debris cloud occur. The 60° incidence achieves the most effective energy dissipation and the highest efficiency owing to the lower normal component of the kinetic energy. The kinetic-energy and velocity–time trajectories of the debris cloud, together with the experimentally observed damage morphologies of the lattice, jointly indicate a transition in the debrisation and collision mode from a fluid-like hypervelocity impact regime to a high-speed penetration regime. The dominant energy-dissipation mechanism in the protective plates shifts from phase transformation and brittle fracture in the initial layer to plastic-deformation energy absorption in subsequent layers. This explains the characteristic decline in energy-dissipation efficiency and is consistent with the experimentally observed increase in flange length from layer to layer in the lattice structure.

By combining numerical simulations with experimental results, the ballistic limit performance of the lattice structure was determined. Compared with traditional aluminum alloy Whipple shields, the lattice structure achieves a total thickness reduction of approximately 25%, while still maintaining excellent load-bearing capacity. In addition, for impact velocities exceeding 3 km/s, the maximum projectile diameter that can be withstood increases on average by 29.1%. This research demonstrates that the proposed trajectory-planning lattice protective structure exhibits outstanding engineering applicability for spacecraft protection against micrometeoroids and space debris. Its reduced thickness and enhanced load-bearing capacity make it a promising candidate for high-value satellite applications.